\documentclass[letterpaper,twocolumn,10pt]{article}
\usepackage{usenix2019_v3}

\usepackage{tikz}
\usepackage{amsmath}
\usepackage{bbm}
\usepackage{amsfonts}
\usepackage{booktabs,multirow}
\usepackage{graphicx,subcaption}
\usepackage{caption,currfile}
\usepackage{enumitem}
\usepackage{amsthm}

\newtheorem{lemma}{Lemma}

\begin{document}

\date{}

\title{\Large \bf An Efficient and Privacy-Preserving Architecture \\for Cross-Institutional Collaborative RAG}

\author{
  {\rm Chenxin Mao$^\dagger$, Shangyu Liu$^\dagger$, Zhenzhe Zheng$^\dagger$, Fan Wu$^\dagger$, Jie Wu$^\ddagger$, Guihai Chen$^\dagger$}\\
  \medskip
  $^\dagger$Shanghai Jiao Tong University \qquad $^\ddagger$Cloud Computing Research Institute, China Telecom
}

\maketitle

\begin{abstract}
Retrieval-Augmented Generation (RAG) empowers LLMs with external knowledge, making cross-institutional domain-specific knowledge base integration a highly promising deployment paradigm. Despite this potential, strict privacy regulations create severe "data silos" that obstruct such collaboration. Building federated RAG systems requires distributed inference, but the Transformer's self-attention mechanism fundamentally conflicts with this by mandating cross-node access to distributed Key-Value caches. To address this challenge, we present FedRAG, a high-throughput, privacy-preserving federated RAG framework. At its core is a novel Scrambled Distributed Attention protocol that utilizes numerically stable feature scrambling and token permutation. By dynamically delegating scrambled computations to collaborating nodes, our system successfully decouples attention execution from data localization without exposing plaintext. Crucially, our approach requires no specialized hardware or model retraining, circumventing the prohibitive latency and communication overheads of cryptographic solutions while robustly defending against intermediate state inversion attacks. Extensive evaluations demonstrate our framework preserves negligible (<0.1\%) model utility degradation and achieves up to a 62$\times$ latency reduction over existing secure baselines, sustaining practical, human-reading throughput for cross-institutional knowledge synergy.
\end{abstract}

\section{Introduction}

In recent years, transformer-based Large Language Models (LLMs) have achieved unprecedented breakthroughs across both academia and industry, demonstrating exceptional capabilities in multilingual interaction, text generation, and logical reasoning~\cite{zhao2023survey}. This paradigm shift has led to their seamless integration into a myriad of applications, significantly optimizing various workflows, such as software development~\cite{chen2021evaluating}, conversational agents~\cite{ouyang2022training}, knowledge management~\cite{pan2024unifying}, and healthcare~\cite{singhal2023large}. It is foreseeable that as their foundational capabilities continue to evolve, LLMs will progressively penetrate more complex, multi-domain applications.

Despite possessing vast pre-trained knowledge bases and exhibiting remarkable generalization, generalized LLMs encounter significant challenges when tasked with domain-specific or highly time-sensitive knowledge. A prominent issue is "hallucination," whereby models, lacking adequate context, are prone to generating plausible yet factually incorrect information~\cite{zhang2025siren}. Furthermore, maintaining the currency of the model's knowledge via continual training entails prohibitive computational and temporal costs. To mitigate these inherent limitations, the Retrieval-Augmented Generation (RAG) architecture has emerged as a promising paradigm~\cite{lewis2020retrieval}. By dynamically integrating external, authoritative knowledge bases, RAG endows LLMs with the ability to access specialized and current information without requiring parameter retraining. This approach effectively curtails the incidence of hallucinations while simultaneously ensuring the accuracy and timeliness of the generated content.

Beyond open-domain applications that leverage publicly accessible web data, many organizations increasingly rely on their proprietary data to construct specialized RAG systems. However, the efficacy of such RAG systems is intrinsically contingent upon the scale and comprehensiveness of their underlying knowledge bases~\cite{borgeaud2022improving}. In highly specialized domains, a single organization typically possesses limited and fragmented proprietary data, which is insufficient to support robust, well-rounded LLM reasoning. Consequently, there is a compelling need to collaborate and synthesize knowledge across multiple institutions to construct a more capable and comprehensive RAG system. For example, healthcare providers can enhance complex diagnostics by integrating electronic medical records from multiple hospitals, financial institutions can conduct robust risk assessments utilizing transaction logs across various banks, and the manufacturing sector can optimize production processes by analyzing operational data from multiple factories. Despite a strong mutual willingness to collaborate, organizations are strictly precluded from directly sharing raw documents due to stringent privacy regulations and the imperative to protect commercial secrets, which inevitably leads to the formation of "data silos." 

Therefore, it is imperative to explore and establish privacy-preserving Federated RAG frameworks to dismantle these informational barriers and enable secure, cross-institutional knowledge synergy. However, developing such collaborative systems entails addressing three primary challenges:

$\bullet$ \textbf{Privacy Protection.} Early works attempted to preserve privacy by transmitting model intermediate states rather than raw text. However, recent studies reveal that the intermediate states of LLMs exhibit high semantic sparsity, rendering them vulnerable to inversion attacks capable of reconstructing partial original text~\cite{dong2025depth}. For instance, Vocabulary Mapping Attacks (VMA)~\cite{thomas2025hidden}, which rely on traversing the vocabulary, can effectively bypass simple permutation-based protection schemes on intermediate states, including both sequence-dimension and hidden-dimension permutations. The emergence of such sophisticated attacks imposes significantly more stringent requirements on the design of privacy-preserving algorithms.

$\bullet$ \textbf{Inference Overhead.} Existing privacy-preserving distributed LLM inference architectures based on cryptographic methods suffer from prohibitive inference overhead. Even state-of-the-art frameworks utilizing Secure Multi-Party Computation (MPC), such as PUMA~\cite{dong2025puma}, require several minutes to generate a single token using an 8-billion parameter model. Furthermore, excessive communication overhead remains a pressing issue. Taking Homomorphic Encryption (HE) schemes (e.g., Paillier~\cite{paillier1999public} or CKKS~\cite{cheon2017homomorphic}) as an example, the ciphertext volume typically expands by tens to hundreds of times compared to the plaintext. For massive intermediate state tensors, such transmission costs are practically unacceptable. The severe latency induced by both computation and transmission renders cryptography-based secure inference impractical for real-world RAG systems.

$\bullet$ \textbf{Universality and Rapid Deployment.} Many distributed secure inference architectures impose stringent prerequisites on the execution environment or the model itself. Hardware-based solutions relying on Trusted Execution Environments (TEEs)~\cite{sabt2015trusted} necessitate specialized hardware support and complex environmental attestation configurations, which are often difficult to satisfy in heterogeneous distributed systems. Conversely, other secure inference architectures require retraining the model~\cite{pang2024bolt,knott2021crypten} or substituting specific operators with approximate counterparts. However, retraining hinders the seamless adoption of the latest high-performance models, while approximate operators inevitably lead to a degradation in inference accuracy.

To address these challenges, we propose FedRAG, a high-throughput, privacy-preserving federated RAG framework designed specifically for cross-institutional knowledge synergy. At its core, FedRAG introduces a novel Scrambled Distributed Attention protocol. Instead of transmitting highly vulnerable intermediate states or relying on heavyweight cryptographic protocols, participating institutions dynamically negotiate a numerically stable feature scrambling mechanism. This specially designed scrambling structure provides robust privacy guarantees while sustaining highly efficient and precise computation on hardware accelerators. By transmitting these scrambled tensors over the network to dynamically delegate computations to an independent compute node, FedRAG successfully decouples the self-attention execution from data localization. This innovative design ensures that no single entity can access the plaintext contexts of others, while fully leveraging the abundant computational resources and high-speed interconnects typical of institutional consortiums.

Our core contributions can be summarized as follows:

$\bullet$ We present FedRAG, a collaborative generative system that breaks down cross-institutional data silos. Our framework requires no model retraining, relies on no approximate operators, and eliminates the need for specialized hardware (e.g., TEEs), enabling seamless and rapid deployment of state-of-the-art LLMs.

$\bullet$ We design a highly efficient, non-cryptographic privacy-preserving attention protocol. It provides robust mathematical and empirical defense against sophisticated intermediate state inversion attacks while replacing prohibitive cryptographic overhead with hardware-friendly linear transformations.

$\bullet$ Extensive evaluations across various open-source LLMs and diverse benchmarks validate the promise of our system. FedRAG achieves up to a 62$\times$ latency reduction over existing secure baselines, sustaining practical, human-reading throughput. Crucially, this immense efficiency gain comes with near-zero model utility degradation across six question answering and summarization benchmarks.

\section{Background}

\subsection{Transformers and Self-Attention}\label{sec:back-transformer}

The Transformer~\cite{vaswani2017attention} serves as the foundational architecture underlying contemporary LLMs, representing a highly efficient and high-performance neural network architecture for processing long sequences. A typical Transformer model is composed of a stack of layers, with each layer primarily consisting of a multi-head self-attention mechanism and a feed-forward network. At the core of this multi-head mechanism is the scaled dot-product attention. During computation, the model first linearly projects each token's hidden state into three matrices\footnote{Throughout this paper, vectors are represented as row vectors, and $d$ denotes the dimension of the attention head.}: Queries ($Q$), Keys ($K$), and Values ($V$). Since $Q$, $K$, and $V$ are all linearly projected from the same input sequence, this mechanism is referred to as \textit{self}-attention, and the attention output is computed as:
\begin{equation}
\text{Attention}(Q, K, V) = \text{softmax}\left(QK^\top/\sqrt{d}\right)V. \label{eq:attention}
\end{equation}

While this dense cross-token interaction enables the model to effectively understand long contexts, it inherently introduces a strict data dependency: the inference for any individual token dictates mandatory access to the representations (i.e., $K$ and $V$) of all other relevant tokens in the sequence. This intrinsic architectural trait forms a fundamental bottleneck when attempting to scale LLM inference across distributed or privacy-sensitive data silos, making the decoupling of attention computation and data localization a critical prerequisite for collaborative generation.

\subsection{Retrieval-Augmented Generation}\label{sec:back-rag}

RAG~\cite{lewis2020retrieval} equips LLMs with external knowledge to mitigate the poor performance and hallucinations caused by static internal weights on domain-specific or time-sensitive tasks, all without requiring costly retraining. A standard query-based RAG pipeline typically executes in three steps. First, an embedding model maps the query into a dense vector space to perform coarse-grained document retrieval via approximate nearest neighbor search. Next, a cross-encoder reranker conducts precision filtering by evaluating fine-grained semantic interactions between the query and candidate documents. Finally, during response synthesis, the top-ranked documents are concatenated with the user query and jointly fed into the LLM to generate a grounded answer.

However, deploying this pipeline in federated scenarios introduces a fundamental architectural conflict. When knowledge bases are distributed across mutually distrustful institutions, centralizing retrieved documents violates strict data privacy constraints. This data fragmentation severely impairs the reranking and generation stages, both of which mandate intensive cross-token attention computation between the user query and the retrieved documents to capture fine-grained semantic dependencies. Because the Transformer's self-attention strictly demands access to all context tokens, executing this computation across distributed, plaintext-sensitive contexts naturally transforms into a challenging cross-node privacy problem, motivating the federated RAG pipeline formalized in \S\ref{sec:motivation}.

\section{System Model and Problem Formulation}
\label{sec:motivation}

In this section, we delineate the unique characteristics of our target deployment scenario, formalize the federated RAG pipeline, and define the threat model that guides our system design.

\subsection{System Model and Target Scenario}

While RAG significantly enhances the performance of LLMs in specialized domains, organizations possessing vast amounts of domain-specific data often face severe ``data silo'' problems due to privacy concerns. Bridging this gap, our work targets a highly practical and increasingly prevalent deployment scenario: \textbf{cross-institutional collaborative RAG within a consortium}. In this scenario, institutions are willing to share information to leverage LLMs for generating macroscopic insights, but they must strictly avoid exposing granular, record-level details to other participating entities. Consequently, we define the primary privacy objective of this work as preventing leakage of the documents' raw plaintext.

In such a cross-institutional setting, the participating nodes are large organizations rather than individual users. They typically operate within enterprise-grade infrastructure characterized by three main features:

$\bullet$ \textbf{Powerful Hardware Resources.} Each participating node operates enterprise-grade servers equipped with high-end AI accelerators and massive storage systems, which provide the computational capacity required to execute large-scale LLM inference.

$\bullet$ \textbf{High-Speed and Predictable Interconnects.} These institutions are often interconnected via dedicated optical networks or co-located within Data Center Colocation facilities, which provide high-speed connections of multi-Gbps and stable millisecond-level low latency.

$\bullet$ \textbf{Multi-Node Participation.} The system inherently involves three or more independent participating nodes. According to the threat model introduced later, we assume no collusion occurs between any pair of these nodes.

Existing privacy-preserving LLM inference frameworks predominantly focus on the traditional ``edge-cloud'' paradigm, which typically involves a resource-constrained edge device communicating over high-latency public networks~\cite{li2025collaborative,zheng2025review}. As a result, most existing work overlooks the critical need for inter-institutional federated inference. To fill this void, our method fully leverages the aforementioned advantages in system resources and multi-participant architectures, proposing a privacy-preserving federated RAG framework designed to satisfy the high-throughput and long-context demands of modern LLMs.

\subsection{Problem Formulation}

To elucidate the system architecture, we formally define the problem of Federated RAG. Consider a consortium of $M$ institutions, where each institution $i$ maintains a private, localized document corpus $\mathcal{D}_i$. The global conceptual corpus is the union of these isolated silos: $\mathcal{D} = \bigcup_{i=1}^{M} \mathcal{D}_i$.

When an authorized user submits a query $q$, the federated RAG pipeline aims to generate a comprehensive response $y$ conditioned on both $q$ and the most relevant documents distributed across $\mathcal{D}$. The end-to-end pipeline can be broadly abstracted into two primary stages: global retrieval and distributed generation. During the \textit{global retrieval} stage, the system collaboratively retrieves and aggregates the most relevant documents across the global conceptual corpus $\mathcal{D}$ to form a globally optimal context pool:
\begin{equation}
C_{final} = \text{Retrieve}\left(q, {\mathcal{D}_1, \mathcal{D}_2, \dots, \mathcal{D}_M}\right).
\end{equation}

Subsequently, in the \textit{distributed generation} stage, an LLM synthesizes the final response. For each generation step $t$, the model computes the attention outputs over the distributed context $C_{final}$ and the query $q$, generating the next token $y_t$ based on the previously generated tokens $y_{<t}$:
\begin{equation}
y_t = \text{LLM}(C_{final}, q, y_{<t}).
\end{equation}

\textbf{Problem Statement:} During the distributed generation (and collaborative re-ranking, if applied), the core operation is the cross-node attention computation $\text{Attention}(Q, K, V)$. The fundamental problem our system addresses is how to compute this cross-node attention collaboratively when $Q$ and $K, V$ belong to mutually distrustful nodes. The system must ensure that neither party can reconstruct the other's exact plaintext tensors, all while maintaining a strict end-to-end latency constraint suitable for real-time applications.

\subsection{Threat Model and Security Assumptions}
\label{sec:threat_model}
Given the consortium-based scenario, our system operates under a highly practical threat model tailored for cross-institutional collaborations.

\textbf{Honest-but-Curious (Semi-Honest) Adversaries:} We assume all participating nodes are \textit{honest-but-curious}. Nodes will strictly adhere to the predefined cryptographic protocols, correct execution of neural network forward passes, and timely transmission of intermediate states. However, they are ``curious'' and may attempt to passively log the exchanged intermediate activations (e.g., scrambled $Q$, $K$, $V$, and attention outputs) to infer or reconstruct the original private text of other institutions. Our primary defensive objective is to thwart such passive reconstruction attacks, notably intermediate state inversion~\cite{dong2025depth} and Vocabulary Mapping Attacks (VMA)~\cite{thomas2025hidden}.

\textbf{Reputation-Aware and Non-Collusion Constraints:} Unlike anonymous P2P networks, institutions in our scenario (e.g., hospitals, banks) are bound by stringent legal frameworks (e.g., HIPAA, GDPR) and commercial contracts. Therefore, we introduce a \textit{Reputation-Aware} assumption: participants are highly risk-averse regarding their institutional reputation. An adversary will actively refrain from any malicious behavior (such as deliberately feeding poisoned data, intentionally altering model weights, or violating protocol rules) if such actions have a high probability of leaving verifiable traces subject to future auditing. Active attacks are thus considered out of scope. Furthermore, this reputation-aware nature inherently enforces a \textit{Non-Collusion} assumption among participating institutions. Any attempt to collude---where two organizations explicitly coordinate to share cryptographic keys or intermediate tensors to decrypt a third party's data---inevitably leaves severe digital footprints (e.g., abnormal out-of-band communication records). More importantly, such malicious coordination exposes each colluding party to the catastrophic risk of being reported or audited by the other. Thus, rational, reputation-bound institutions will inherently refrain from collusion.

\section{Scrambled Distributed Attention}\label{sec:algo_basis}

\subsection{Distributed Attention Computation}\label{sec:dist_attn}

In distributed RAG systems,  the KV Cache of the LLM is partitioned across multiple nodes, requiring cross-node collaboration to execute the attention mechanism. To minimize cross-node communication, we build on the online-softmax decomposition widely adopted in efficient attention implementations~\cite{dao2022flashattention,liu2023ring}, which decomposes attention into two stages: per-node local computation followed by a lightweight global aggregation. In what follows, we formalize this decomposition in our federated setting; this framework will serve as the substrate on which our privacy-preserving protocol operates  (\S4.4). 

\textbf{Per-Node Local Computation.} Assume that the relevant KV Cache is distributed across $M$ nodes. Let $K_{i}$ and $V_{i}$ denote the key and value matrices stored on node $i$, respectively. Given a query matrix $Q$, node $i$ directly applies the standard attention computation to its local $(K_{i}, V_{i})$ producing two quantities:

$\bullet$ \textbf{Node-Local Attention Output.} Applying Eq.~\eqref{eq:attention} to the locally available $K$ and $V$ matrices gives
\begin{equation}
    O_{i} = \text{softmax}\left(\frac{QK_{i}^\top}{\sqrt{d}}\right)V_{i}.
\end{equation}
This output is normalized only over the keys stored on node $i$ and thus does not yet reflect the global attention distribution.

$\bullet$ \textbf{Local Normalization Factor.} To recover the global attention distribution across all nodes, we also compute the corresponding normalization factor. Let $\mathbf{w}_{i} \in \mathbb{R}^{L_Q}$ denote the vector whose $\ell$-th entry aggregates the unnormalized softmax weights on node $i$:
\begin{equation}
    [\mathbf{w}_{i}]_\ell = \sum\nolimits_{t=1}^{L_{V_{i}}} \exp\left(\frac{QK_{i}^\top}{\sqrt{d}}\right)_{\ell,t}.
\end{equation}

\textbf{Global Aggregation.} Finally, the local results from all nodes are communicated and aggregated to reconstruct the global attention output. The global attention output is reconstructed as
\begin{equation}
    \text{Attention}(Q, K, V) =
    \frac{\sum\nolimits_{i=1}^M \mathbf{w}_{i} \odot \mathbf{O}_{i}}{\sum\nolimits_{i=1}^M \mathbf{w}_{i}},
\end{equation}
where $\odot$ and the division are applied row-wise.

This two-stage decomposition decouples the attention computation across nodes while preserving mathematical equivalence to the centralized formulation. Crucially, it exposes a clean interface—each node only exchanges ($O_i$, $\mathbf{w}_i$) with the aggregator—on which we layer our privacy-preserving transformation in \S4.4.

\subsection{Privacy-Preserving Attention Computation}

\textbf{Random Feature Scrambling.} Let $\Phi \in \mathbb{R}^{d \times d}$ be an invertible matrix,  which we call a \emph{scrambling matrix}. Given a vector $x \in \mathbb{R}^d$, we refer to $x\Phi$ or $x\Phi^{-\top}$ as a \emph{scrambled vector}; the choice between $\Phi$ and  $\Phi^{-\top}$ depends on the role of $x$ in the downstream  attention computation, as will become clear in \S4.4. The KV Cache of the LLM is partitioned across multiple nodes, requiring cross-node collaboration to execute the attention mechanism. We defer the concrete construction of $\Phi$ to the next paragraph, since the following two algebraic properties hold for any invertible $\Phi$.

\begin{lemma}[Inner Product Preservation]\label{lemma:inner_product}
For any two vectors $x_1, x_2 \in \mathbb{R}^d$ and any invertible matrix $\Phi \in \mathbb{R}^{d \times d}$, scrambling with the dual matrices  $\Phi$ and $\Phi^{-\top}$ preserves the inner product:
\begin{equation}
    \langle x_1 \Phi,\, x_2 \Phi^{-\top} \rangle
    = (x_1 \Phi)(x_2 \Phi^{-\top})^\top
    = \langle x_1, x_2 \rangle.
\end{equation}
\end{lemma}

\begin{lemma}[Linearity]\label{lemma:linearity}
For any vectors $x_1, \ldots, x_n \in \mathbb{R}^d$ scrambled with the same matrix $\Phi \in \mathbb{R}^{d \times d}$, and any scalar weights $w_1, \ldots, w_n \in \mathbb{R}$:
\begin{equation}
    \sum_{i=1}\nolimits^n w_i (x_i \Phi) 
    = \left( \sum_{i=1}\nolimits^n w_i x_i \right) \Phi.
\end{equation}
\end{lemma}

These two properties are chosen by design: they correspond precisely to the two core operations of scaled dot-product attention. Lemma \ref{lemma:inner_product} preserves the Q-K inner products that determine attention logits, while Lemma \ref{lemma:linearity} preserves the V-weighted summation that produces attention outputs. Together, they enable a computation party to execute attention on scrambled $(Q,K,V)$ and return a result that, after output-side descrambling, equals the plaintext attention—without ever observing the plaintext tensors. We formalize this protocol in \S4.4.

\textbf{Constructing a Numerically Stable Scrambling Matrix.} 
While Lemmas \ref{lemma:inner_product} and \ref{lemma:linearity} hold for any invertible $\Phi$, a naive choice—such as a fully random dense matrix sampled from a continuous distribution—fails in practice. Modern LLM inference runs in low-precision floating-point formats (BF16 or FP16), which offer only 7-10 bits of mantissa. A random dense $\Phi$ produces entries of widely varying magnitudes, causing the matrix-vector products $x\Phi$ and $x\Phi^{-\top}$ to suffer from severe round-off and cancellation errors. These errors accumulate across the dozens of transformer layers, degrading generation quality to the point of unusability (please refer to Appendix~\ref{sec:appendix-numerical-stability} for an empirical demonstration). A practical scrambling matrix must therefore reconcile two seemingly conflicting goals: sufficient mixing to obscure the plaintext, and numerical stability under low-precision arithmetic.

\noindent\textbf{Our construction.}
We construct the scrambling matrix as:
\begin{equation}
\Phi = S_1 P_1 H P_2 S_2,
\label{eq:phi-construction}
\end{equation}
where $H \in \mathbb{R}^{d \times d}$ is a normalized Hadamard matrix,  $P_1, P_2$ are independent random permutation matrices, and $S_1, S_2$ are random diagonal matrices whose entries possess random signs and variable magnitudes. Each component serves a distinct role:

\begin{itemize}[wide, nosep]
\item \textbf{Normalized Hadamard matrix $H$} simultaneously addresses both goals from above. Its base $\{-1, +1\}$ entries allow matrix multiplication to be computed via stable sign-flipping additions followed by a single global $1/\sqrt{d}$ scaling, avoiding the floating-point errors of arbitrary dense multiplications (\emph{numerical stability}). At the same time, its orthogonality ensures that each output coordinate is a $\pm 1/\sqrt{d}$-weighted sum of all input coordinates (\emph{sufficient mixing}).
\item \textbf{Permutation matrices $P_1, P_2$} randomize the otherwise deterministic $H$. A fixed $H$ alone is publicly known; sandwiching it between two random permutations yields a distribution over mixing matrices. Permutations are exact operations, so they preserve numerical stability.
\item \textbf{Scaling matrices $S_1, S_2$} distort distances. Pure permutation-and-Hadamard preserves $L_2$ distances between rows, exposing topological information (e.g., kNN structure) that enables graph-matching attacks (\S\ref{sec:privacy}). The random scaling factors severely disrupt these distance relationships. Their magnitudes are sampled from an empirical range (e.g., $[1/8, 8]$), representing a fundamental trade-off between privacy enhancement and floating-point precision.
\end{itemize}

The two-layer sandwich structure ($S_1 P_1 \cdot H \cdot P_2 S_2$) ensures that both the input and output sides of $H$ are randomized, so partial knowledge of one layer does not help an attacker invert the other.

\noindent\textbf{Security under structured $\Phi$.}
A natural concern is whether this structured $\Phi$---which has fewer effective degrees of freedom than a fully random matrix---provides sufficient privacy. In \S\ref{sec:privacy}, we show that it does: the highly collinear geometry of LLM hidden states~\cite{ethayarajh2019contextual, gao2021simcse} fundamentally limits any algebraic inversion attack, making the structural constraints on $\Phi$ largely inconsequential in practice. We defer the detailed analysis to \S\ref{sec:privacy}.

\textbf{Random Token Permutation.} In addition to scrambling the feature dimension, we introduce random permutation in the token dimension. Due to the permutation-equivariant nature of attention computation, applying a consistent permutation to the $K$ and $V$ matrices yields an attention output from which the original result can be exactly recovered. Similarly, permuting the $Q$ matrix results in an attention output that, upon applying the inverse permutation, matches the original output. This token-level shuffling significantly enhances system resilience against known-plaintext attacks and brute-force enumeration.

\textbf{Scrambled Attention Protocol.} Consider three matrices $Q = [q_1; \dots; q_{L_Q}] \in \mathbb{R}^{L_Q \times d}$, $K = [k_1; \dots; k_{L_K}] \in \mathbb{R}^{L_K \times d}$, and $V = [v_1; \dots; v_{L_V}] \in \mathbb{R}^{L_V \times d}$, where $L_V = L_K$. We generate two random scrambling matrices $\Phi_{KQ}, \Phi_{V} \in \mathbb{R}^{d \times d}$ and two permutation matrices $P_Q \in \mathbb{R}^{L_Q \times L_Q}, P_{KV} \in \mathbb{R}^{L_V \times L_V}$ to form the key set $\Theta$. The workflow is as follows:
\begin{enumerate}[topsep=1pt, itemsep=0pt, parsep=0pt, partopsep=0pt]
\setlength{\abovedisplayskip}{3pt}
\item Plaintext Matrix Scrambling:
\begin{align}
    \text{Enc}(Q,K,V,\Theta) &= (P_Q Q \Phi_{KQ}, P_{KV} K \Phi_{KQ}^{-\top}, P_{KV} V \Phi_{V}) \notag \\
    & = (Q', K', V').
\end{align}
\item Scrambled Attention Computation:
\begin{align}
    \text{At}&\text{tention}(Q', K', V') \notag = \text{softmax}\left(Q' {K'}^\top/\sqrt{d}\right) V' \notag \\
    &= \text{softmax}\left(P_Q Q \Phi_{KQ} (P_{KV} K \Phi_{KQ}^{-\top})^\top/\sqrt{d}\right) (P_{KV} V \Phi_{V}) \notag \\
    &= P_Q \text{softmax}\left(QK^\top/\sqrt{d}\right) P_{KV}^\top (P_{KV} V \Phi_{V}) \notag \\
    &= P_Q \text{Attention}(Q, K, V) \Phi_{V}.
\end{align}
\item Output Descrambling:
\begin{align}
    \text{Dec}&(\text{Attention}(Q', K', V'),\Theta) \notag \\
    &= P_Q^{-1} \text{Attention}(Q', K', V') \Phi_{V}^{-1} \notag \\
    &= \text{Attention}(Q, K, V).
\end{align}
\end{enumerate}

In this design, no single node can access the plaintext data of matrices from other nodes, yet the system correctly executes the attention computation.

\section{The RAG Pipeline}

\begin{figure}[t]
\centering
\includegraphics[width=0.9\linewidth]{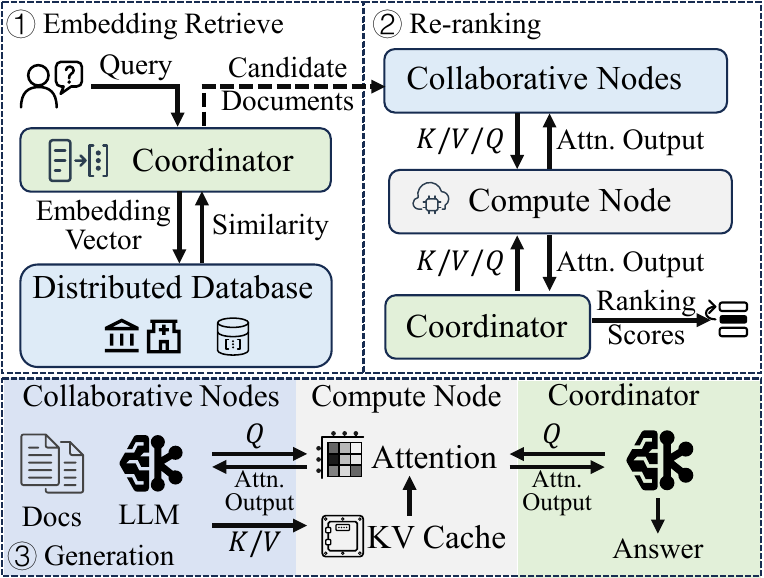}
\caption{System architecture of the federated RAG system.}
\label{fig:system}
\end{figure}

This section details the lifecycle of a single user request within the federated RAG system. When an authorized user from a participating organization requires generative services, they submit a query to their local node. This receiving node assumes the role of the \textit{Coordinator} for the request, orchestrating subsequent operations alongside other collaborating nodes in the federation system. 

\subsection{System Overview and Node Roles}
A complete RAG request lifecycle consists of three main phases: vector retrieval, re-ranking, and generation. The system architecture is illustrated in Figure~\ref{fig:system}. In the vector retrieval phase, the  \textit{Coordinator} first converts the received user query into a vector representation using an embedding model and broadcasts it to all collaborating nodes. Each node performs vector retrieval against its local documents, returning the identifiers and similarity scores of the top-$k$ most relevant documents to the  \textit{Coordinator}. The  \textit{Coordinator} then aggregates and ranks these scores to select the top-$k$ global candidates for the subsequent re-ranking stage. During re-ranking, the  \textit{Coordinator} and the collaborating nodes jointly execute the re-ranking model inference to determine the final top-$m$ ($m<k$) documents that constitute the context for the RAG request. Finally, in the generation phase, the  \textit{Coordinator} collaboratively performs LLM inference with other nodes holding the retrieved contexts to generate the final response. 

Since both the re-ranking and generation phases require cross-node attention computation, we first formalize the roles of participating nodes. To execute the privacy-preserving attention mechanism across the federated network, we categorize the nodes into three distinct roles based on their functions during each cross-node attention computation:

$\bullet$ \textbf{Inquirer.} The node currently running the forward pass for unprocessed tokens. It possesses the plaintext Query required for the attention computation.

$\bullet$ \textbf{Context Owner.} The node possessing the plaintext $K$ and $V$ matrices required for the attention computation.

$\bullet$ \textbf{Compute Node.} The node responsible for executing the actual attention computation. It receives scrambled matrices from the Inquirer and Context Owner, computes the result, and returns it to the Inquirer.

For each cross-node attention computation, the Inquirer and Context Owner share the scrambling key set $\Theta$. The Compute Node does not possess these keys and thus cannot reconstruct the information, ensuring the security of the privacy-preserving protocol. These roles are applied dynamically across both the re-ranking and generation stages.

\subsection{Collaborative Re-ranking}

Building upon the vector retrieval results, the re-ranking stage employs specialized re-ranking models (e.g., BGE Reranker~\cite{chen2024bge}, mGTE~\cite{zhang2024mgte}) to recalibrate the relevance scores between candidate documents and the query. Unlike the bi-encoder architecture used in the preceding vector retrieval phase, which independently maps queries and documents into a shared vector space to compute coarse-grained similarities, this stage adopts a cross-encoder approach. It directly models the fine-grained, token-level interactions between each query-document pair, thereby significantly improving retrieval quality.

\textbf{Input Construction.} In this stage, the model input is a concatenation of the query and the candidate document: $\texttt{[CLS]} \oplus \text{query} \oplus \texttt{[SEP]} \oplus \text{document} \oplus \texttt{[SEP]}$, where $\texttt{[CLS]}$ and $\texttt{[SEP]}$ are special tokens functioning as the global sequence representation and the segment separator, respectively. In a federated RAG system, the query and documents may reside on different nodes, necessitating cross-node attention computation. We utilize the scrambling-based privacy-preserving attention method introduced in \S\ref{sec:algo_basis} to execute collaborative re-ranker inference without leaking private information.

\textbf{Privacy-Preserving Forward Pass.} Most classical re-ranking models typically adopt an encoder-only architecture utilizing bidirectional attention~\cite{xu2025survey,yates2021pretrained}. This implies that the attention computation for every single token requires full access to the intermediate states of the entire sequence (i.e., both preceding and succeeding contexts). Consequently, participating nodes simultaneously act as both Inquirers (seeking attention outputs for their own tokens) and Context Owners (providing their $K$ and $V$ matrices to others) during the forward pass of each layer. Nodes independently compute the $Q$, $K$, and $V$ matrices for their local tokens based on the output of the previous layer. These matrices are scrambled and transmitted to the designated Compute Node responsible for the attention calculation. Once the computation is complete, this node sends the corresponding attention outputs back to the respective Inquirer nodes owning the tokens. The participating nodes then descramble the results and proceed with the subsequent Feed-Forward Network (FFN) computations to complete the layer's inference. This process repeats until all layers are processed, requiring bidirectional communication with the Compute Node $L_{\text{reranker}}$ times (where $L_{\text{reranker}}$ is the number of model layers). Finally, the hidden state of the \texttt{[CLS]} token is mapped to a relevance score.

\subsection{Collaborative Generation}

\begin{figure}[t]
\centering
\includegraphics[width=0.95\linewidth]{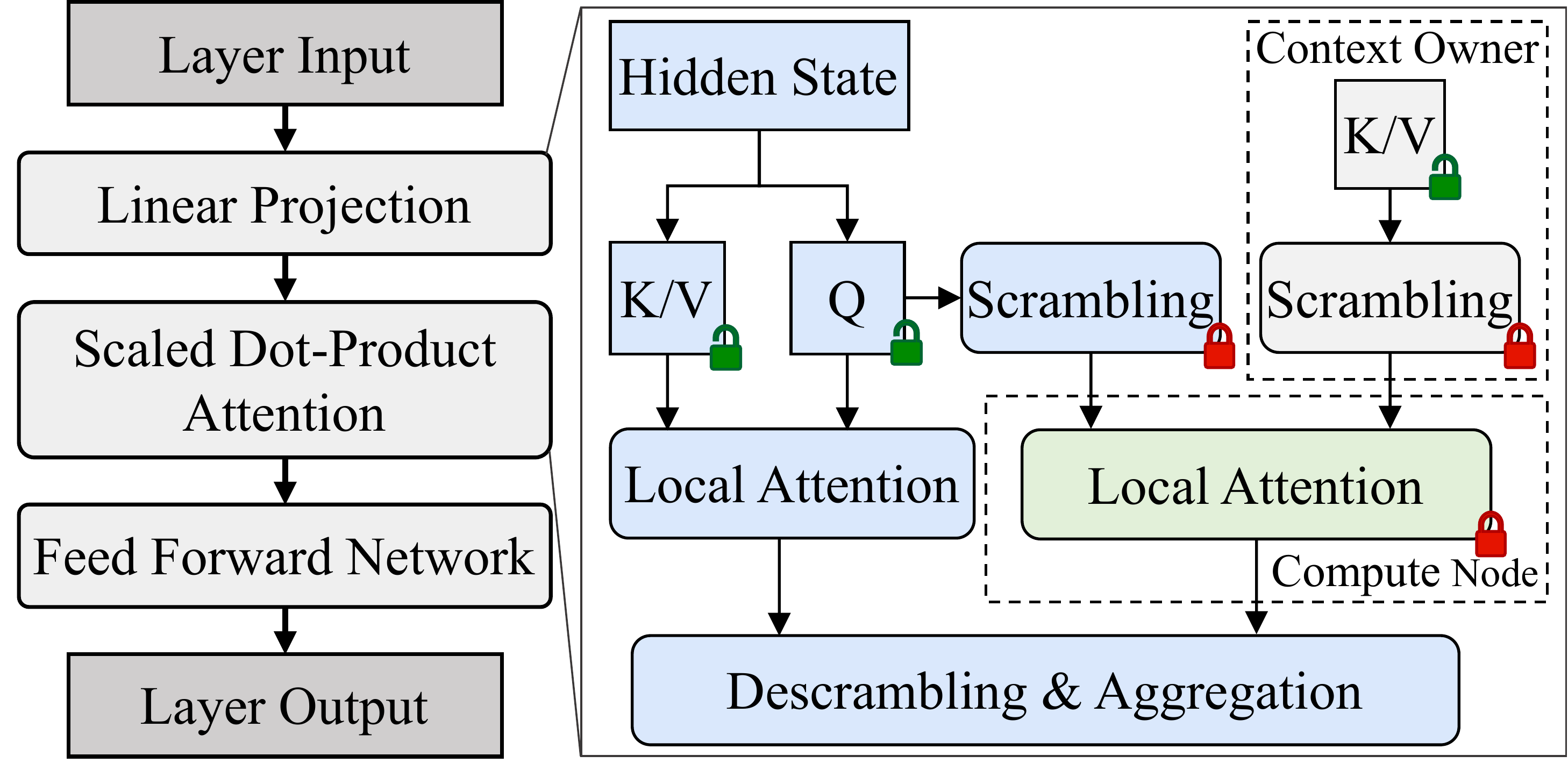}
\caption{Forward pass of a decoder layer and the distributed attention computation in the generation phase.}
\label{fig:decoder-layer}
\end{figure}

In the RAG generation phase, the retrieved results are fed into the LLM along with the user input. Modern LLMs predominantly rely on a decoder-only architecture. In the forward pass of a decoder-only transformer, each token only computes attention with its preceding tokens, rather than all tokens in the sequence as is characteristic of encoder architectures. During the generation process, the intermediate state (i.e., the KV Cache) of each token remains unchanged once computed. For distributed generation in our system, this implies that the roles of nodes participating in the decoder's attention computation are asymmetric: a Context Owner does not need to perform layer-by-layer attention computation synchronously with the Inquirer possessing subsequent tokens. Instead, it can complete the KV Cache computation for its local tokens in advance and transmit them to the Compute Node ahead of time. We illustrate the forward pass of a single decoder layer and our distributed attention computation in Figure~\ref{fig:decoder-layer}.

The generation process comprises a prefill phase and a decoding phase. The former builds the KV Cache for the input tokens, while the latter generates new tokens step-by-step based on the prefilled cache. Unlike centralized RAG systems, the prefill and decoding attention computations in a federated RAG system must be coordinated across nodes due to the distributed nature of documents and queries. We define a continuous sequence of plaintext tokens held by a single node as an \textit{input segment}. The global input sequence consists of ordered input segments $T_1, T_2, \dots, T_{n_s}$, owned by nodes $\mathcal{O}(T_1), \mathcal{O}(T_2), \dots, \mathcal{O}(T_{n_s})$, respectively. Nodes owning these segments sequentially perform the forward pass to complete attention caching based on the segment's position in the sequence.

\textbf{Prefill Phase.} When the tokens in segment $T_i$ undergo the forward pass, the owners of all preceding segments, $\mathcal{O}(T_1), \dots, \mathcal{O}(T_{i-1})$, act as Context Owners. They scramble the $K$ and $V$ of their already-prefilled tokens and transmit them to the Compute Node in advance. Concurrently, the owner of $T_i$ assumes the role of the Inquirer. For each transformer layer, the Inquirer computes the local $Q$, $K$, and $V$ matrices for its current segment and performs a partitioned attention calculation. As mathematically formulated in \S\ref{sec:dist_attn}, it first derives the local attention output utilizing its own $K$, $V$, and $Q$ matrices. Next, the Inquirer scrambles its local Query matrix and dispatches it to the Compute Node. The Compute Node executes the attention computation between this scrambled Query and the scrambled $K$, $V$ previously received from the Context Owners, and subsequently returns the result to the Inquirer. Upon receiving this result, the Inquirer descrambles it and aggregates it with the previously computed local attention output to finalize the attention computation for the current segment. The aggregated attention output is then fed into the subsequent FFN to complete the forward pass for that specific layer. This layer-wise process is iterated until the forward computation for all layers is concluded. Finally, during transmission idle periods, the Inquirer scrambles the $K$ and $V$ matrices of its newly processed segment and forwards them to the Compute Node, making them available as context for subsequent prefill and decoding stages.

\textbf{Decoding Phase.} This phase commences after the last input segment completes prefill. Since the final input segment exactly corresponds to the user's initial query, its owner, $\mathcal{O}(T_{n_s})$, is inherently the \textit{Coordinator} of the RAG request. Having already assumed the Inquirer role during the prefill of this final segment, the \textit{Coordinator} naturally retains this identity to commence generating new tokens, while the owners of all preceding segments serve as Context Owners. The decoding forward pass mirrors the prefill phase, with the distinction that it processes one newly generated token at a time to predict the next token. Furthermore, the Inquirer does not need to send the newly computed $K$ and $V$ of the generated token to the Compute Node immediately. For each layer, the Inquirer scrambles the new token's Query vector, sends it to the Compute Node (which already holds the scrambled KV Cache from the prefill phase), receives the result, descrambles it, and proceeds with the FFN. Upon completing the final decoder layer, the Inquirer samples the new token and initiates the next generation step. This process repeats until a termination condition is met (e.g., reaching a maximum length limit or generating an end-of-sequence token).

\subsection{Dynamic Role Assignment}\label{sec:method-role}

Our privacy-preserving attention scheme employs a symmetric-like scrambling mechanism, implying that any node capable of scrambling is also capable of descrambling. This necessitates a strict constraint: a node acting as a Compute Node in a specific attention computation must not simultaneously be a Context Owner or an Inquirer. Otherwise, the Compute Node could use the shared key to reverse the counterparty's data, violating privacy. This constraint implies that our federated RAG system requires a minimum of three independent physical nodes.

If there exists a node in the system that does not own any input segments, it can permanently serve as the Compute Node for all cross-node attention computations in that request. However, in scenarios with few nodes and a uniform document distribution, all nodes may own input segments. This requires dynamically assigning the Compute Node role to at least two different physical nodes during a request. When physical node $\mathcal{N}^E_1$ acts as a Context Owner, $\mathcal{N}^E_2$ must serve as the Compute Node, and vice versa. This implies that the KV Cache for certain segments must be replicated to both $\mathcal{N}^E_1$ and $\mathcal{N}^E_2$, incurring additional communication overhead.

Let $\text{maxSeg}(\mathcal{N}_i) = \max_{\mathcal{O}(T_j)=\mathcal{N}_i} j$ denote the highest index of an input segment owned by physical node $\mathcal{N}_i$. Consider two physical nodes assigned the Compute Node role, $\mathcal{N}^E_1$ and $\mathcal{N}^E_2$, with $\text{maxSeg}(\mathcal{N}^E_1) < \text{maxSeg}(\mathcal{N}^E_2)$. To minimize the overhead of redundant transmission, we strategically select the node with the smallest maximum segment index as one of the Compute Nodes. Specifically, if there exists a physical node $\mathcal{N}_i$ such that $\text{maxSeg}(\mathcal{N}_i) = 1$ (i.e., it owns only the very first input segment), we can permanently assign it as a Compute Node. This strategy eliminates the need to redundantly transmit the $K$ and $V$ information of any input segment.

\section{Privacy Analysis}\label{sec:privacy}

\textbf{State Inversion \& Text Recovery.} State inversion attacks aim to recover original tokens by accessing the model's intermediate activation states. Representative methods include VMA~\cite{thomas2025hidden}, white-box optimization-based inversion~\cite{dong2025depth,erdougan2022unsplit}, and black-box generation-based inversion~\cite{dong2025depth, morris2023text}. VMA relies on traversing the vocabulary and utilizing the sorted $L_1$-distance for ANN collisions. However, the dense mixing introduced by the Hadamard matrix $H$ and the non-isometric scaling by $S_1, S_2$ in our scrambling scheme thoroughly destroy the distance relationships between the plaintext and ciphertext, rendering VMA ineffective. White-box optimization-based inversion requires calculating the loss via forward propagation to construct a differentiable distance function; since the computational node lacks the scrambling matrix $\Phi$, backpropagation cannot be performed. Furthermore, black-box generation-based inversion relies on a stationary feature space to train a Seq2Seq inverse mapping model. In contrast, our system dynamically negotiates $\Phi$ for each layer at the beginning of every RAG request, preventing the attacker from obtaining a stable feature space for model convergence. Consequently, to execute the aforementioned text recovery attacks, an adversary must satisfy an absolute prerequisite: unmixing the dense scrambled representation manifold to extract the plaintext intermediate states. This forces the attacker to resort to low-level algebraic unmixing attacks.

\begin{figure}[t]
\centering
\begin{subfigure}[b]{0.49\columnwidth}
\centering
\includegraphics[width=\linewidth]{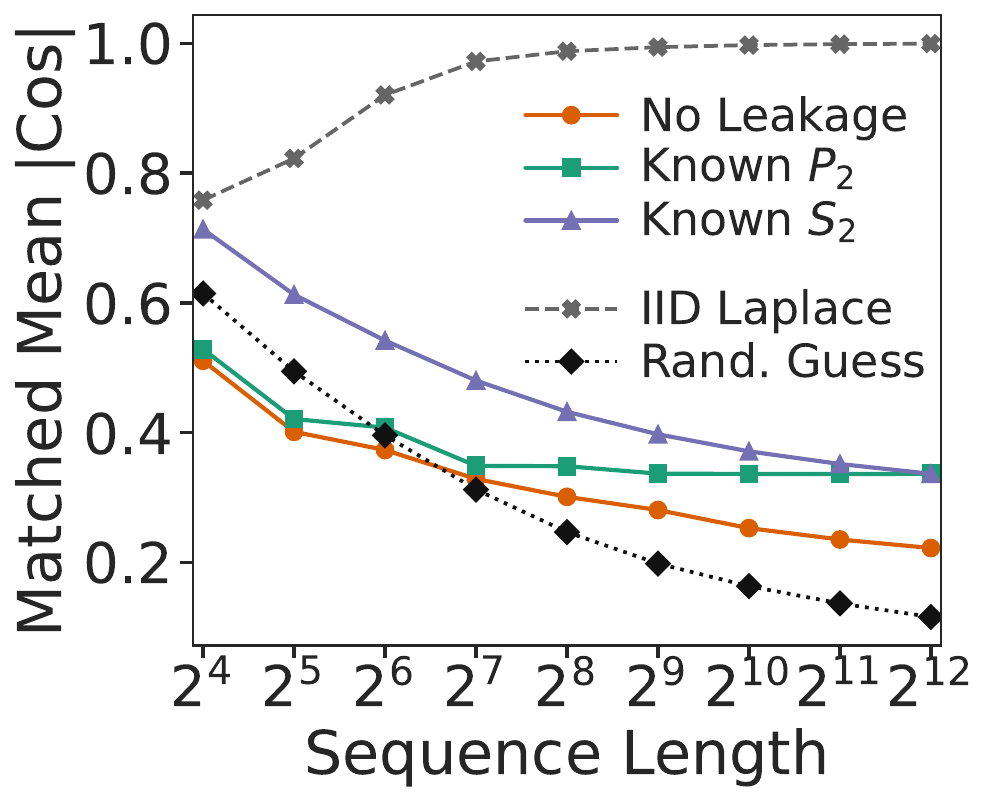}
\caption{Resistance to ICA.}
\label{fig:security-ica}
\end{subfigure}
\hfill
\begin{subfigure}[b]{0.49\columnwidth}
\centering
\includegraphics[width=\linewidth]{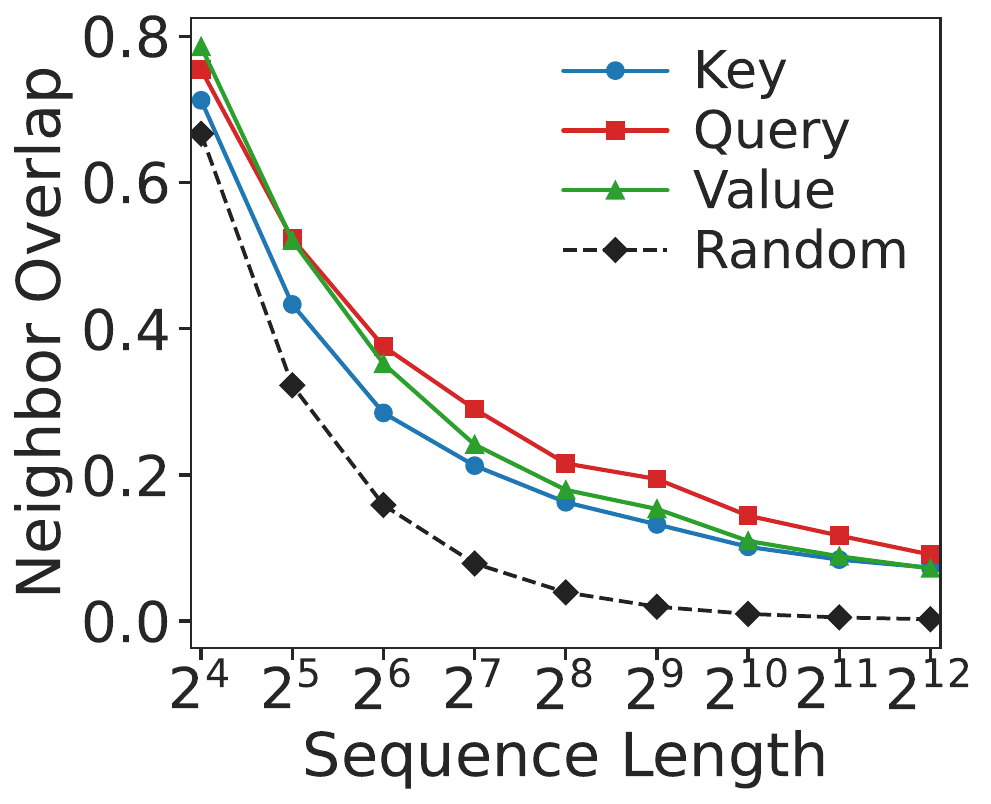}
\caption{Resistance to graph matching.}
\label{fig:security-knn_overlap}
\end{subfigure}
\caption{Security evaluation of the scrambling mechanism against algebraic and topological attacks.}
\label{fig:security-eval}
\end{figure}

\textbf{Defeating Independent Component Analysis (ICA).} ICA is a classic blind source separation algorithm that aims to extract original underlying signals from their linear mixtures without prior knowledge of the mixing parameters~\cite{hyvarinen2000independent}. We evaluated the resistance of our method against ICA attacks, as shown in Figure~\ref{fig:security-ica}. The five curves represent the Hungarian Mean Absolute Cosine Similarity between the attack results and the targets under different scenarios: no leakage (knowing only the scrambling structure), known $P_2$, known $S_2$, a random guess baseline based on the empirical distribution of intermediate states, and a positive control using independent Laplace sources. These results demonstrate that even under pessimistic assumptions of partial information leakage, the recovery similarities plateau near the random guess baseline and fail to approach an exploitable degree. In contrast, the IID Laplace positive control can be almost perfectly recovered. This disparity demonstrates that our scheme naturally resists ICA as the highly collinear intermediate states (i.e., the anisotropy problem) of modern LLMs intrinsically violate the independent source assumption~\cite{ethayarajh2019contextual, gao2021simcse} upon which the algorithm relies.

\textbf{Graph Matching Attacks.} Graph matching attacks are non-parametric attack schemes that attempt to reconstruct features by aligning the relative distance matrices of the scrambled space and the known plaintext spaces. Since exact subgraph matching is fundamentally an NP-Hard problem~\cite{conte2004thirty}, existing efficient graph matching algorithms heavily rely on the topological stability of local spatial features~\cite{sogaard2018limitations}. As illustrated in Figure~\ref{fig:security-knn_overlap}, experiments demonstrate that the kNN overlap rate between the plaintext and ciphertext spaces drops rapidly as the sequence length $L$ increases. At $L=4096$, the local neighborhood overlap for $Q$, $K$, and $V$ all plummet below 15\%, which is vastly below the theoretical threshold that existing efficient alignment methods can exploit~\cite{cullina2016improved}.

\textbf{Malicious Prompt.} To counter the threat of request initiators attempting to steal other nodes' documents via maliciously crafted prompts, our system employs a game-theoretic defense mechanism based on ciphertext auditing. The initiating node of a RAG request (i.e., the Coordinator) must send the scrambled query to the computational nodes, and its keys are shared with at least one other node in the system. In the event of a dispute, the federation can collaboratively audit these scrambled records. This ensures proactive suppression of malicious attacks under a ``reputation-aware'' mechanism.

\section{Evaluation}
\subsection{Evaluation Setup}
We implement our privacy-preserving mechanism as an API-compatible, drop-in replacement for the standard attention function in the Hugging Face transformers library. We conduct our evaluations on a server equipped with one NVIDIA PRO 6000 and two NVIDIA RTX 5090 GPUs. To enforce inter-node isolation, we deploy the participating nodes within separate Docker containers, and we employ Toxiproxy as the network proxy between containers to simulate diverse network environments with varying latency and bandwidth constraints.

In this section, we first evaluate the end-to-end latency, throughput, and communication volume of our method compared to existing privacy-preserving LLM inference techniques in RAG scenarios. Subsequently, we analyze the impact of various factors—such as the number of protected layers, network latency, and network bandwidth—on system efficiency. Next, we benchmark the generation quality of our method across six diverse datasets to demonstrate that our scrambling mechanism incurs negligible degradation in model accuracy. Finally, we investigate the accuracy-efficiency trade-offs of quantizing intermediate states to provide insights for future optimizations.

\textbf{Baselines.} We select SCX~\cite{yuan2025scx} and PermLLM~\cite{zheng2024permllm} as our primary baselines, as they share similar characteristics with our approach: they do not require model retraining, do not mandate TEEs, and avoid approximate operators. For SCX, we implement it according to the non-TEE environment settings described in their paper. For PermLLM, to ensure a fair comparison, we strictly apply its cryptographic operations solely to the attention computation.

\textbf{Models.} We evaluate our system using multiple state-of-the-art open-source LLMs, including the Qwen 3~\cite{yang2025qwen3}, Llama 3.1~\cite{grattafiori2024llama}, Ministral 3~\cite{liu2026ministral}, and GPT-OSS series~\cite{agarwal2025gpt}, with parameter sizes ranging from 4B to 20B. All models are obtained from the Hugging Face model hub and executed using their default data types. For the RAG retrieval phase, we uniformly employ bge-m3 and bge-reranker-v2-m3~\cite{chen2024bge} as the embedding and reranking models, respectively. During the reranking stage, we apply 4-bit quantization to the intermediate states, which we discuss in further detail in \S\ref{sec:evaluation-quantization}.

\textbf{Datasets.} For throughput and latency evaluations, we generate input token sequences of fixed lengths, with configurations detailed in Table \ref{tab:length-configs}. Our experimental setup distributes context tokens across two nodes, each hosting two equal-length documents. For the reranking phase, we use 10 candidate documents for recalibration.

For accuracy evaluation, we select several prominent question-answering (QA) and summarization datasets equipped with reference contexts. These benchmarks cover a diverse range of domains, context lengths, and reasoning complexities. Specifically, we evaluate our system on five standard QA datasets—SQuAD~\cite{rajpurkar2016squad}, HotpotQA~\cite{yang2018hotpotqa}, MuSiQue~\cite{trivedi2022musique}, MS MARCO~\cite{bajaj2016ms}, and AmbigQA~\cite{min2020ambigqa}—along with one long-form meeting summarization benchmark, QMSum~\cite{zhong2021qmsum}. Detailed descriptions of these datasets are deferred to Appendix~\ref{sec:appendix-datasets}.

\begin{table}[t]
\centering
\caption{Length configurations in throughput evaluations.}
\label{tab:length-configs}
\begin{tabular}{cccc}
\hline
Config & Context & Query & Output \\
\hline
Short & 256+256 & 32 & 32 \\
Medium & 1024+1024 & 64 & 64 \\
Long & 4096+4096 & 128 & 128 \\
\hline
\end{tabular}
\end{table}

\subsection{End-to-End Latency and Communication Overhead}
\begingroup
\begin{figure*}[t]
\centering
\includegraphics[width=\textwidth]{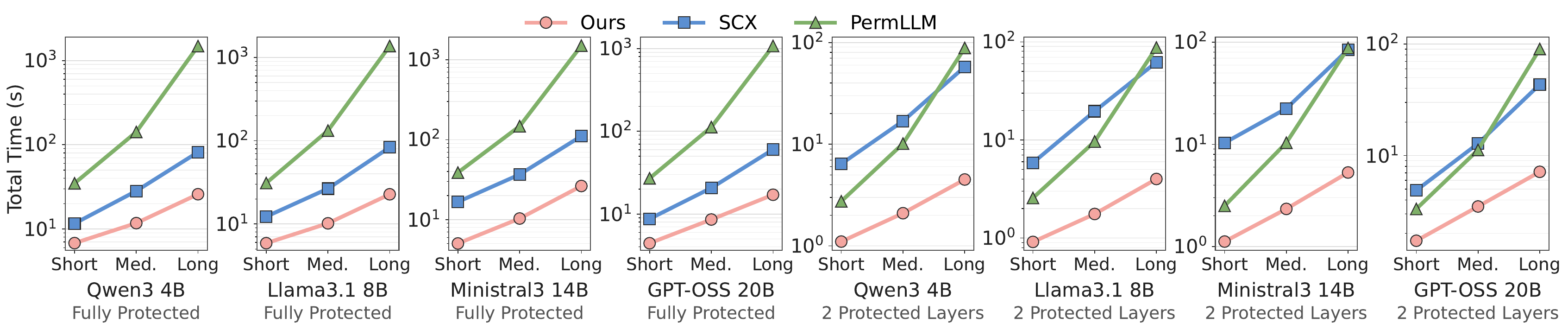}
\caption{End-to-end latency of RAG requests.}
\label{fig:overall}
\end{figure*}
\endgroup

\begin{table*}[t]
\centering
\small
\setlength{\tabcolsep}{4pt}
\caption{Performance evaluation for fully protected generation.}
\label{tab:overall-details-full}
\resizebox{\textwidth}{!}{
\begin{tabular}{llcccccccc}
\toprule
\multirow{2}{*}{Length} & \multirow{2}{*}{Method} & \multicolumn{4}{c}{\texttt{meta-llama/Llama-3.1-8B-Instruct}} & \multicolumn{4}{c}{\texttt{openai/gpt-oss-20b}} \\
\cmidrule(lr){3-6} \cmidrule(lr){7-10}
 & & TTFT (s) $\downarrow$ & Dec. TPS $\uparrow$ & Traff. (MiB) $\downarrow$ & Comm. Rounds $\downarrow$ & TTFT (s) $\downarrow$ & Dec. TPS $\uparrow$ & Traff. (MiB) $\downarrow$ & Comm. Rounds $\downarrow$ \\
\midrule
\multirow{3}{*}{Short} & Ours & \textbf{3.23} & \textbf{11.70} & \textbf{301.17} & \textbf{1147} & \textbf{1.39} & \textbf{10.19} & \textbf{220.66} & \textbf{867} \\
 & PermLLM & 11.93 & 1.62 & 4721.73 & 10851 & 9.68 & 1.79 & 4224.67 & 8163 \\
 & SCX & 6.22 & 5.13 & 1791.25 & 3458 & 3.52 & 5.94 & 1181.29 & 2618 \\
\midrule
\multirow{3}{*}{Medium} & Ours & \textbf{4.83} & \textbf{11.79} & \textbf{1107.61} & \textbf{2171} & \textbf{2.45} & \textbf{10.28} & \textbf{804.73} & \textbf{1635} \\
 & PermLLM & 89.23 & 1.46 & 45815.71 & 21091 & 73.47 & 1.62 & 38419.13 & 15843 \\
 & SCX & 15.09 & 5.48 & 6752.25 & 6530 & 10.13 & 5.98 & 4454.75 & 4922 \\
\midrule
\multirow{3}{*}{Long} & Ours & \textbf{12.00} & \textbf{11.85} & \textbf{4238.54} & \textbf{4219} & \textbf{7.72} & \textbf{13.54} & \textbf{3037.73} & \textbf{1647} \\
 & PermLLM & 1243.09 & 0.96 & 587393.74 & 41571 & 966.51 & 1.16 & 483843.42 & 31203 \\
 & SCX & 60.15 & 5.50 & 26202.25 & 12674 & 39.13 & 6.01 & 17274.69 & 9530 \\
\bottomrule
\end{tabular}
}
\end{table*}

Figure \ref{fig:overall} illustrates the total end-to-end latency of a single RAG request for our method and the baselines across varying prompt lengths. Although prior studies have demonstrated that transmitting plaintext intermediate states from any single layer can lead to privacy leakage (as discussed in \S\ref{sec:privacy}), we follow the partial-layer protection strategy adopted by SCX and introduce an additional configuration that only protects two layers for comparison, with results presented in the second row of Figure \ref{fig:overall}. In practice, this partial-layer protection strategy serves as a viable trade-off option between system efficiency and privacy, allowing users to configure the system according to their specific operational requirements.

As shown in Figure \ref{fig:overall}, our method achieves the lowest end-to-end latency across all six configurations and four models, yielding a 1.71$\times$ to 62.96$\times$ speedup over the baselines. Table \ref{tab:overall-details-full} details the comprehensive efficiency metrics for the Llama-3.1-8B model, with results for other models provided in Appendix \ref{sec:appendix-overall-details}. The table highlights that our approach achieves significantly superior results in Time to First Token (TTFT), Decode TPS, communication rounds, and communication volume. Crucially, the overhead of our method exhibits a linear growth trend with respect to sequence length, endowing our framework with excellent scalability for increasingly prevalent long-context workloads. We attribute this substantial efficiency improvement to the core design principle that we maximize the computational payload delegated to the Compute Node within a single cross-node communication round, and we rely exclusively on hardware-efficient matrix multiplications for both plaintext encryption and ciphertext computation, thereby minimizing latency induced by cryptography and communication.

\subsection{Throughput and Network Sensitivity}\label{sec:evaluation-throughput}
\begingroup
\graphicspath{{\currfiledir}}
\begin{figure*}[t]
\captionsetup{skip=2pt}
\centering
\includegraphics[width=\textwidth]{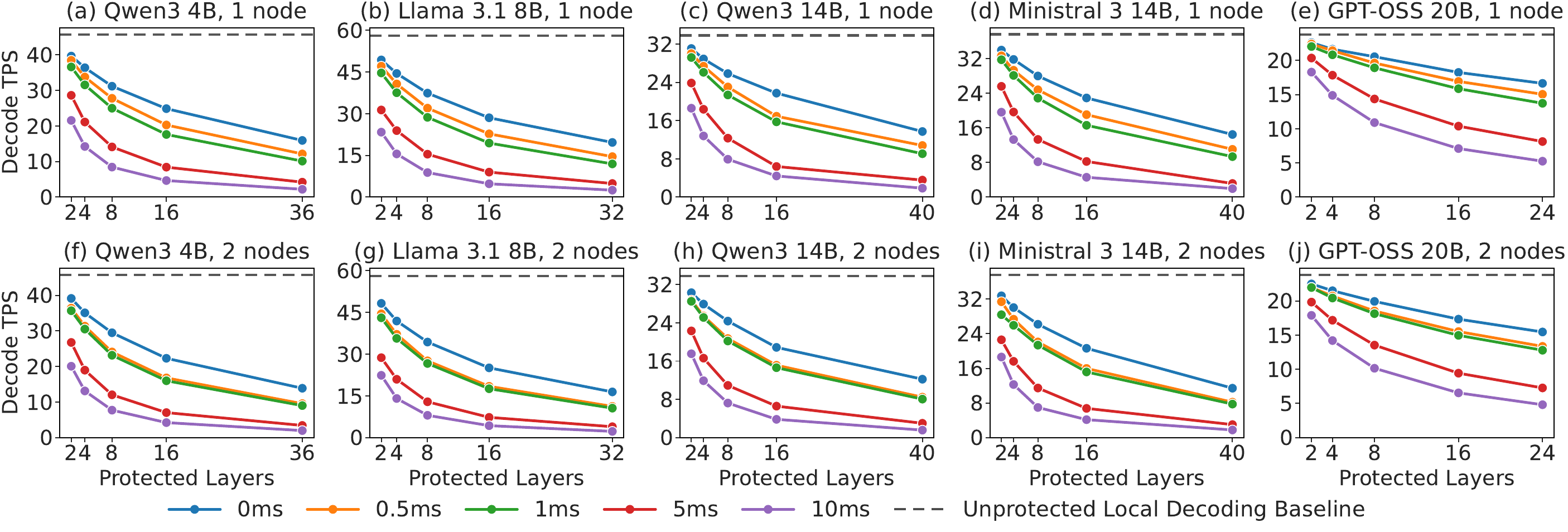}
\caption{Decode throughput under different protected-layer counts and network settings.}
\label{fig:ours-decode-throughput-grid-full}
\end{figure*}
\endgroup

\begingroup
\graphicspath{{\currfiledir}}
\newlength{\BandwidthPanelHeight}
\setlength{\BandwidthPanelHeight}{0.3\columnwidth}
\begin{figure}[t]
\captionsetup{skip=2pt}
\captionsetup[subfigure]{justification=centering, singlelinecheck=false, skip=2.0pt}
\centering
\includegraphics[width=0.99\columnwidth]{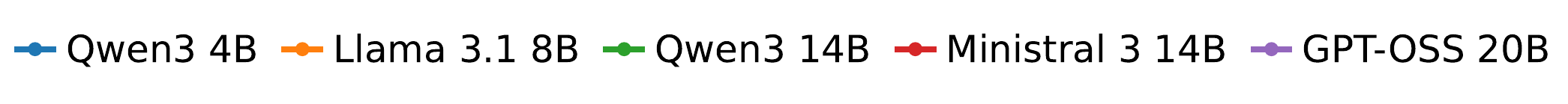}
\par\vspace{-0.4em}
\begin{subfigure}[t]{0.49\columnwidth}
\centering
\includegraphics[height=\BandwidthPanelHeight]{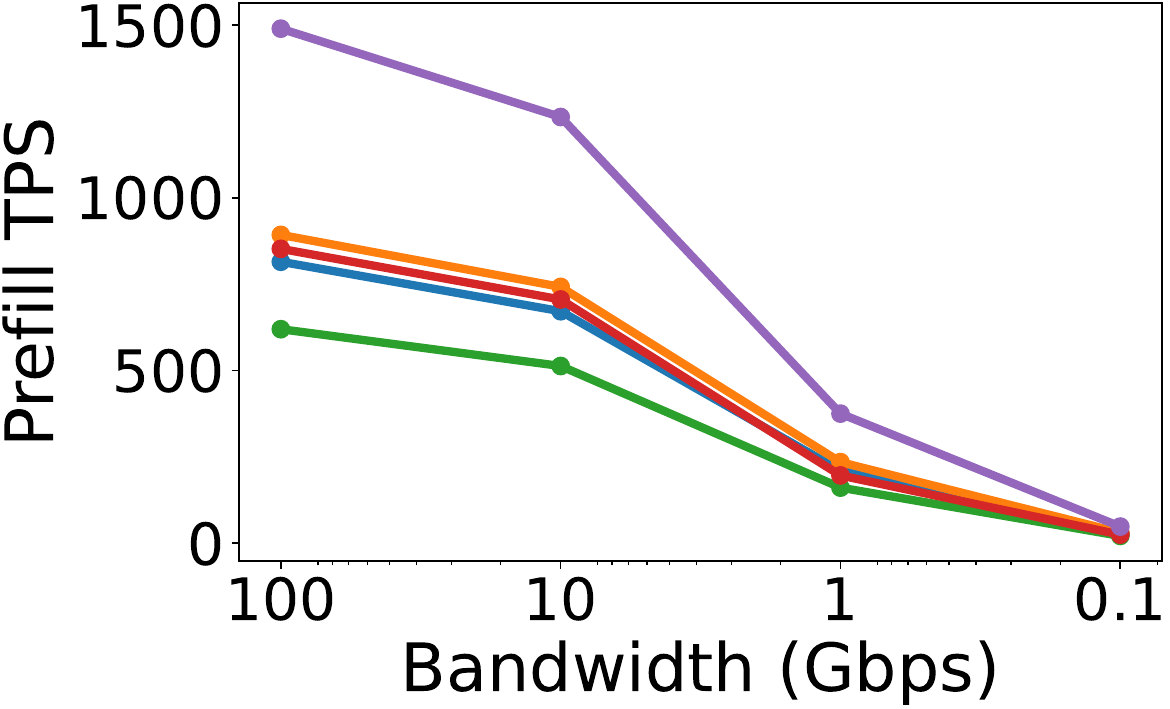}
\caption{Prefill TPS.}
\end{subfigure}\hfill
\begin{subfigure}[t]{0.49\columnwidth}
\centering
\includegraphics[height=\BandwidthPanelHeight]{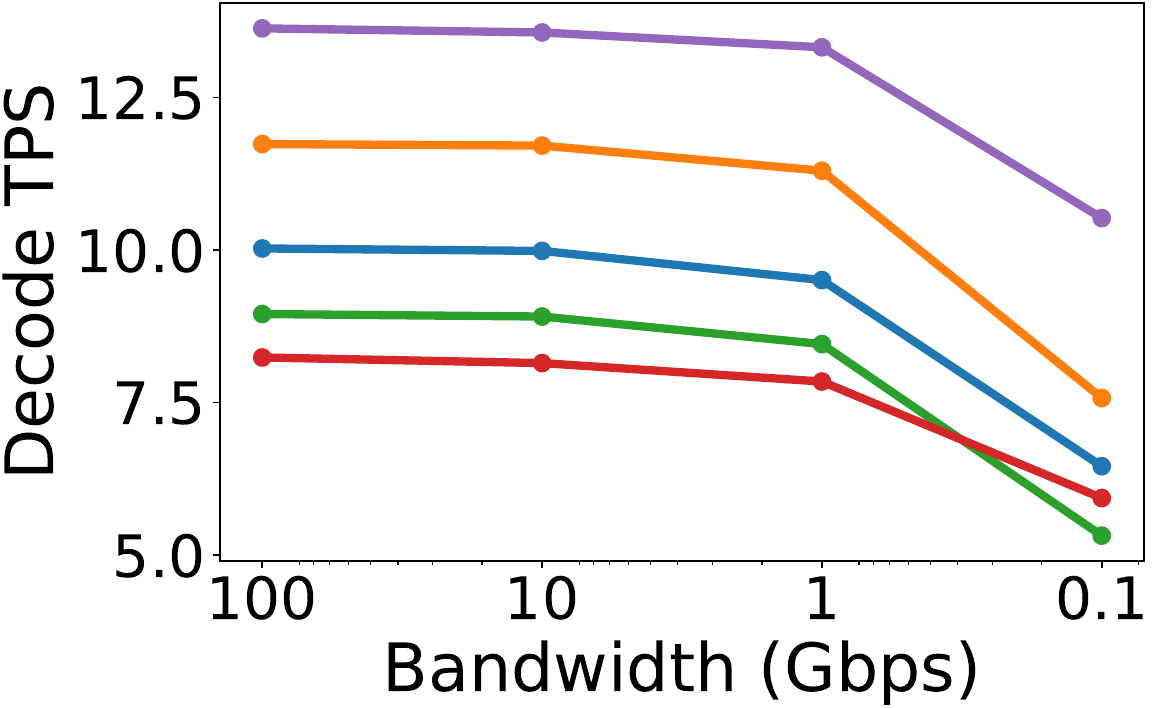}
\caption{Decode TPS.}
\end{subfigure}
\caption{Throughput vs. bandwidth.}
\label{fig:ours-bandwidth-prefill-decode}
\end{figure}
\endgroup

Figure \ref{fig:ours-decode-throughput-grid-full} plots the Decode TPS metrics under varying network latencies. Here, the number of nodes denotes the physical machines that act as Compute Nodes during a given RAG request; as articulated in \S\ref{sec:method-role}, if a request requires all nodes in the system to supply tokens, we must dynamically assign the Compute Node role to two distinct nodes to guarantee privacy protection. The empirical results demonstrate that with a Round-Trip Time (RTT) of approximately 1ms, our method sustains Decode TPS greater than 10 under the full-layer protection configuration. For collaborative RAG scenarios deployed across physical servers within the same data center, across availability zones in the same region, or across departments within a research institution, our system can deliver output speeds matching human reading rates while providing uncompromising privacy~\cite{kwon2025vllm}. When the RTT increases to 10ms, the Decode TPS drops to single digits; in such scenarios, operators can resort to offline generation or intentionally trade a portion of privacy for enhanced throughput. Nevertheless, our throughput remains significantly higher than existing cryptographic solutions, which typically require several minutes to generate a single token~\cite{dong2025puma}.

The throughput for configurations employing two Compute Nodes is depicted in the second row of Figure \ref{fig:ours-decode-throughput-grid-full}. Compared to the single Compute Node setup, the throughput exhibits an average degradation of roughly 10\%. This decline primarily occurs because, during the decoding phase, the Inquirer must transmit the query vector of the token to the second Compute Node and wait for both nodes to finalize their computations before proceeding to the subsequent inference step.

We further investigate the impact of network bandwidth on throughput, with the results shown in Figure \ref{fig:ours-bandwidth-prefill-decode}. The two subfigures illustrate the throughput trends for the prefill and decoding phases, respectively. We observe that network bandwidth significantly affects the prefill phase, whereas the Decode TPS does not exhibit noticeable degradation until the bandwidth drops to 100Mbps. This discrepancy is due to the fact that the prefill phase involves fewer communication rounds but transmits massive volumes of data per round, whereas the decoding phase is the exact opposite; consequently, network latency serves as the predominant performance bottleneck for decoding.

\subsection{Utility Preservation on Downstream Tasks}\label{sec:evaulation-acc}
\begin{table*}[t]
\centering
\small
\setlength{\tabcolsep}{3pt}
\caption{Impact of Distributed Attention and Feature Scrambling on Downstream Task Performance.}
\label{tab:qa-six-dataset-results-compressed}
\resizebox{\linewidth}{!}{\begin{tabular}{lccc@{}p{6pt}@{}ccc@{}p{6pt}@{}ccc@{}p{6pt}@{}ccc@{}p{6pt}@{}ccc@{}p{6pt}@{}ccc}
\toprule
\multirow{2}{*}{Method} & \multicolumn{3}{c}{SQUAD} &  & \multicolumn{3}{c}{HOTPOTQA} &  & \multicolumn{3}{c}{MARCO} &  & \multicolumn{3}{c}{AMBIGQA} &  & \multicolumn{3}{c}{MUSIQUE} &  & \multicolumn{3}{c}{QMSUM} \\
\cmidrule(lr){2-4} \cmidrule(lr){6-8} \cmidrule(lr){10-12} \cmidrule(lr){14-16} \cmidrule(lr){18-20} \cmidrule(lr){22-24}
 & Acc $\uparrow$ & R-L $\uparrow$ & PPL $\downarrow$ &  & Acc $\uparrow$ & R-L $\uparrow$ & PPL $\downarrow$ &  & Acc $\uparrow$ & R-L $\uparrow$ & PPL $\downarrow$ &  & Acc $\uparrow$ & R-L $\uparrow$ & PPL $\downarrow$ &  & Acc $\uparrow$ & R-L $\uparrow$ & PPL $\downarrow$ &  & F1 $\uparrow$ & R-L $\uparrow$ & PPL $\downarrow$ \\
\midrule
Baseline & \textbf{89.43} & 89.67 & 3.06 &  & \textbf{69.40} & \textbf{78.26} & \textbf{3.07} &  & \textbf{22.16} & \textbf{45.92} & \textbf{4.23} &  & \textbf{49.94} & \textbf{64.02} & 17.0 &  & \textbf{63.93} & \textbf{66.79} & \textbf{3.33} &  & 26.29 & 20.24 & 16.1 \\
+Distributed & 89.42 & \textbf{89.68} & 3.06 &  & 69.35 & 78.24 & 3.07 &  & 22.09 & 45.84 & 4.23 &  & 49.89 & 63.95 & 17.0 &  & 63.82 & 66.66 & 3.33 &  & \textbf{26.55} & \textbf{20.38} & \textbf{16.1} \\
+Scrambler & 89.42 & 89.68 & \textbf{3.06} &  & 69.37 & 78.21 & 3.07 &  & 22.11 & 45.84 & 4.24 &  & 49.79 & 63.91 & \textbf{16.9} &  & 63.79 & 66.67 & 3.34 &  & 26.53 & 20.32 & 16.1 \\
\bottomrule
\end{tabular}
}
\end{table*}

In \S\ref{sec:algo_basis}, we theoretically proved that our privacy-preserving method is mathematically equivalent to the standard attention computation. However, in practical deployments, floating-point arithmetic inevitably introduces numerical precision errors, which can cause deviations in the model's final output. To evaluate the impact of these deviations on model capabilities, we run benchmark evaluations across six datasets. We select three instruction-tuned models from the previous tests and additionally introduce the Qwen2.5-32B-Instruct model; possessing 64 hidden layers, it better exposes the cumulative effects of precision loss during the forward pass. We design ablation baselines for the benchmark, sequentially introducing the distributed attention computation and the intermediate state scrambling mechanism to analyze their respective impacts on model utility.

The experimental results are summarized in Table \ref{tab:qa-six-dataset-results-compressed}. The accuracy reported in the table denotes the proportion of samples where the ground truth answer appears within the model's output; for the meeting summarization dataset QMSum, we use the F1 score instead of accuracy. Furthermore, we measure the ROUGE-L overlap between the generated output and the ground truth, as well as the perplexity (PPL) of the ground truth given the model. The empirical results reveal that we maintain a virtually identical PPL compared to the standard attention baseline, with accuracy and ROUGE-L experiencing negligible average drops of roughly 0.06\% and 0.05\%, respectively. These findings confirm that our precision-preserving scrambling implementation enables privacy protection with near-zero accuracy degradation.

\subsection{Accuracy-Efficiency Trade-offs via Quantization}\label{sec:evaluation-quantization}

\begingroup
\graphicspath{{\currfiledir}{\currfiledir figures/}}
\begin{figure}[t]
\captionsetup[subfigure]{justification=centering, singlelinecheck=false, skip=2.0pt}
\captionsetup{skip=2pt}
\centering
\begin{subfigure}[t]{0.49\linewidth}
\centering
\includegraphics[width=\linewidth]{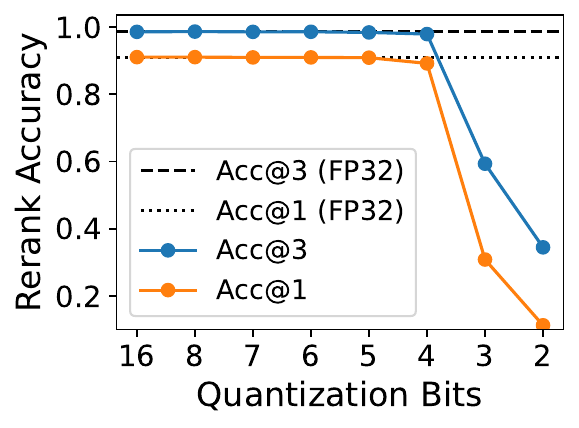}
\caption{Accuracy under quantization.}
\label{fig:retriever-quantization-accuracy}
\end{subfigure}\hfill
\begin{subfigure}[t]{0.49\linewidth}
\centering
\includegraphics[width=\linewidth]{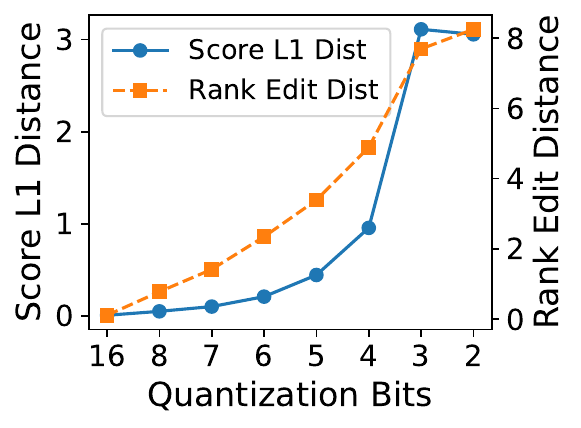}
\caption{Deviation from the FP32 baseline.}
\label{fig:retriever-quantization-deviation}
\end{subfigure}
\caption{Impact of intermediate activation quantization on the Reranker.
}
\label{fig:retriever-quantization-two-panel}
\end{figure}
\endgroup

Quantization serves as a vital optimization to reduce communication overhead, which is particularly prominent during the reranking phase (consuming $\sim$70\% of the total transmission using the model's default precision). To investigate this accuracy-efficiency trade-off, we apply standard affine min-max quantization~\cite{jacob2018quantization} to the intermediate states during reranking. Using the SQuAD dataset, we retrieve the top-10 candidate documents via dense embeddings and re-score them using the quantized reranker.

Figures \ref{fig:retriever-quantization-accuracy} and \ref{fig:retriever-quantization-deviation} illustrate the impact of varying quantization bit-widths on retrieval quality and raw output deviations. While absolute score deviations (L1 and Rank Edit distances) increase progressively below 8-bit precision, the actual reranking accuracy (Acc@1 and Acc@3) remains remarkably robust down to 4-bit quantization, maintaining performance comparable to the FP32 baseline. This robustness stems from the fact that reranking efficacy depends strictly on the relative ordering of scores rather than their absolute logits, providing a sufficiently large margin to absorb quantization errors. At 3-bit and below, however, inconsistencies rapidly spike, rendering the results unusable. Thus, we strongly recommend 4-bit quantization for the reranking phase as the optimal balance between communication efficiency and retrieval quality.

Furthermore, we briefly investigate intermediate state quantization during the generation phase. Directly applying 8-bit quantization to the scrambled states severely degrades generation quality because the two random scaling operations ($S_1$ and $S_2$) significantly expand the tensor's numerical distribution range. For deployments under extremely constrained bandwidth, operators can adopt a workaround by deliberately removing the post-Hadamard scaling matrix ($S_2$). Under this relaxed privacy configuration, 8-bit quantization incurs only marginal utility penalties (average drops of $\sim$0.26\% in accuracy and $\sim$0.23\% in ROUGE-L, as detailed in Appendix \ref{sec:appendix-quantization}), allowing systems to dynamically trade rigorous privacy for communication efficiency.

\section{Discussion}

\textbf{System Scalability.} In a typical RAG pipeline, the number of documents utilized during the reranking and generation phases is bounded by a predefined constant (i.e., top-$k$ and top-$m$ documents). Consequently, the number of participating nodes in these two stages does not scale indefinitely with the overall system size. As the institutional consortium expands, the primary performance bottleneck shifts to the global vector retrieval phase. Fortunately, a substantial body of existing work has extensively investigated efficient distributed vector search in large-scale systems~\cite{wang2021milvus, johnson2019billion}. These established techniques can be orthogonally integrated into our framework to alleviate retrieval bottlenecks.

\textbf{Architectural Compatibility.} Since our privacy-preserving mechanism exclusively modifies the attention computation, it is inherently compatible with a wide spectrum of multimodal models that rely on the standard Transformer architecture. Furthermore, for emerging sub-quadratic architectures (e.g., Mamba~\cite{gu2023mamba}, DeltaNet~\cite{yang2024parallelizing}), our framework remains adaptable. For layers utilizing these alternative mechanisms, the intermediate states can be directly transmitted to the Inquirer node for local computation, while the remaining standard attention layers can continue to leverage our distributed, scrambled attention protocol.

\textbf{Stronger Privacy Guarantees.} While our previous evaluations primarily explored trading a portion of privacy for enhanced throughput under constrained networks, our scrambling methodology can be naturally extended to provide stronger privacy guarantees in scenarios with abundant bandwidth or stringent confidentiality requirements. By deliberately exchanging communication overhead for heightened security, we can apply more aggressive obfuscation to the intermediate states, particularly in shallower layers. Potential extensions include elevating the transmission precision of intermediate states to 32-bit floating-point coupled with a fully random dense scrambling matrix, as well as applying dimensional expansion to the $K$ and $Q$ matrices padded with orthogonal random noise. These enhancements would further resist sophisticated algebraic unmixing attempts.

\textbf{Future Work.} Our work represents an early exploration into privacy-preserving federated RAG systems, leaving several avenues for future research. Currently, our decoding phase processes the generation of a single token sequentially. Future iterations could incorporate advanced mechanisms such as Speculative Decoding~\cite{leviathan2023fast} to generate multiple tokens per cross-node communication round, thereby significantly boosting system throughput. Additionally, while our current defense against malicious prompts relies on reactive ciphertext auditing logs, future designs could integrate automated detection and mitigation mechanisms to proactively block adversarial queries prior to execution.

\section{Related Work}

\textbf{Retrieval-Augmented Generation.} RAG was initially proposed to mitigate LLM hallucinations and bypass costly retraining by fetching relevant, non-parametric knowledge from external data stores prior to generation~\cite{lewis2020retrieval}. Existing RAG architectures can be fundamentally categorized into three paradigms based on their augmentation methodologies~\cite{zhao2026retrieval}. \textit{Query-based RAG} directly concatenates retrieved plaintext information with user inputs at the prompt level (e.g., REALM~\cite{guu2020retrieval}, SELF-RAG~\cite{asai2023self}). \textit{Latent representation-based RAG} deeply fuses retrieved contexts into the generative model's hidden states via cross-attention mechanisms (e.g., FiD~\cite{izacard2021leveraging}, RETRO~\cite{borgeaud2022improving}). Finally, \textit{Logit-based RAG} interpolates retrieval probabilities into the generator's step-wise decoding distribution during inference (e.g., kNN-LM~\cite{khandelwal2019generalization}, TRIME~\cite{zhong2022training}).

\textbf{Privacy-Preserving LLM Inference.} Recent work has actively explored privacy-preserving LLM inference to safeguard data in collaborative deployments. Cryptography-based methods~\cite{hao2022iron, dong2025puma} deliver rigorous mathematical security by executing neural network operations over encrypted data or secret shares, primarily leveraging Fully Homomorphic Encryption (FHE) and Secure Multi-Party Computation (SMPC). Alternatively, TEE-based methods~\cite{luo2025shadow} leverage hardware enclaves to protect critical data and sensitive intermediate states. Furthermore, statistical perturbation mechanisms~\cite{du2023dp, zeng2025privacyrestore} mitigate feature inversion risks while avoiding extensive cryptographic computational overhead by applying Differential Privacy (DP) principles to inject controlled noise into intermediate LLM activations.

\section{Conclusion}

In this paper, we present a privacy-preserving federated RAG framework designed to overcome the prevalent data silo problem and the prohibitive latency constraints of existing cryptographic inference techniques. By leveraging a numerically stable feature scrambling mechanism alongside dynamic role delegation, our system facilitates secure, cross-node attention computation without exposing plaintext contexts. Extensive evaluations confirm that our approach delivers orders of magnitude latency reduction and sustains practical, human-reading throughput with near-zero model utility degradation. Ultimately, this work eliminates critical infrastructure barriers, paving the way for secure, cross-institutional knowledge synergy in highly regulated and privacy-sensitive domains.

\bibliographystyle{plain}
\bibliography{001_reference}

\begin{thebibliography}{10}

\bibitem{agarwal2025gpt}
Sandhini Agarwal, Lama Ahmad, Jason Ai, Sam Altman, Andy Applebaum, Edwin Arbus, Rahul~K Arora, Yu~Bai, Bowen Baker, Haiming Bao, et~al.
\newblock gpt-oss-120b \& gpt-oss-20b model card.
\newblock {\em arXiv preprint arXiv:2508.10925}, 2025.

\bibitem{asai2023self}
Akari Asai, Zeqiu Wu, Yizhong Wang, Avirup Sil, and Hannaneh Hajishirzi.
\newblock Self-rag: Learning to retrieve, generate, and critique through self-reflection.
\newblock In {\em The Twelfth International Conference on Learning Representations}, 2023.

\bibitem{bajaj2016ms}
Payal Bajaj, Daniel Campos, Nick Craswell, Li~Deng, Jianfeng Gao, Xiaodong Liu, Rangan Majumder, Andrew McNamara, Bhaskar Mitra, Tri Nguyen, et~al.
\newblock Ms marco: A human generated machine reading comprehension dataset.
\newblock {\em arXiv preprint arXiv:1611.09268}, 2016.

\bibitem{borgeaud2022improving}
Sebastian Borgeaud, Arthur Mensch, Jordan Hoffmann, Trevor Cai, Eliza Rutherford, Katie Millican, George~Bm Van Den~Driessche, Jean-Baptiste Lespiau, Bogdan Damoc, Aidan Clark, et~al.
\newblock Improving language models by retrieving from trillions of tokens.
\newblock In {\em International conference on machine learning}, pages 2206--2240. PMLR, 2022.

\bibitem{chen2024bge}
Jianlv Chen, Shitao Xiao, Peitian Zhang, Kun Luo, Defu Lian, and Zheng Liu.
\newblock Bge m3-embedding: Multi-lingual, multi-functionality, multi-granularity text embeddings through self-knowledge distillation.
\newblock {\em arXiv preprint arXiv:2402.03216}, 4(5), 2024.

\bibitem{chen2021evaluating}
Mark Chen, Jerry Tworek, Heewoo Jun, Qiming Yuan, Henrique Ponde De~Oliveira Pinto, Jared Kaplan, Harri Edwards, Yuri Burda, Nicholas Joseph, Greg Brockman, et~al.
\newblock Evaluating large language models trained on code.
\newblock {\em arXiv preprint arXiv:2107.03374}, 2021.

\bibitem{cheon2017homomorphic}
Jung~Hee Cheon, Andrey Kim, Miran Kim, and Yongsoo Song.
\newblock Homomorphic encryption for arithmetic of approximate numbers.
\newblock In {\em International conference on the theory and application of cryptology and information security}, pages 409--437. Springer, 2017.

\bibitem{conte2004thirty}
Donatello Conte, Pasquale Foggia, Carlo Sansone, and Mario Vento.
\newblock Thirty years of graph matching in pattern recognition.
\newblock {\em International journal of pattern recognition and artificial intelligence}, 18(03):265--298, 2004.

\bibitem{cullina2016improved}
Daniel Cullina and Negar Kiyavash.
\newblock Improved achievability and converse bounds for erdos-r{\'e}nyi graph matching.
\newblock {\em ACM SIGMETRICS performance evaluation review}, 44(1):63--72, 2016.

\bibitem{dao2022flashattention}
Tri Dao, Dan Fu, Stefano Ermon, Atri Rudra, and Christopher R{\'e}.
\newblock Flashattention: Fast and memory-efficient exact attention with io-awareness.
\newblock {\em Advances in neural information processing systems}, 35:16344--16359, 2022.

\bibitem{dong2025depth}
Tian Dong, Yan Meng, Shaofeng Li, Guoxing Chen, Zhen Liu, and Haojin Zhu.
\newblock Depth gives a false sense of privacy:$\{$LLM$\}$ internal states inversion.
\newblock In {\em 34th USENIX Security Symposium (USENIX Security 25)}, pages 1629--1648, 2025.

\bibitem{dong2025puma}
Ye~Dong, Wen-jie Lu, Yancheng Zheng, Haoqi Wu, Derun Zhao, Jin Tan, Zhicong Huang, Cheng Hong, Tao Wei, Wen-Guang Chen, et~al.
\newblock Puma: Secure inference of llama-7b in five minutes.
\newblock {\em Security and Safety}, 4:2025014, 2025.

\bibitem{du2023dp}
Minxin Du, Xiang Yue, Sherman~SM Chow, Tianhao Wang, Chenyu Huang, and Huan Sun.
\newblock Dp-forward: Fine-tuning and inference on language models with differential privacy in forward pass.
\newblock In {\em Proceedings of the 2023 ACM SIGSAC Conference on Computer and Communications Security}, pages 2665--2679, 2023.

\bibitem{erdougan2022unsplit}
Ege Erdo{\u{g}}an, Alptekin K{\"u}p{\c{c}}{\"u}, and A~Erc{\"u}ment {\c{C}}i{\c{c}}ek.
\newblock Unsplit: Data-oblivious model inversion, model stealing, and label inference attacks against split learning.
\newblock In {\em Proceedings of the 21st Workshop on Privacy in the Electronic Society}, pages 115--124, 2022.

\bibitem{ethayarajh2019contextual}
Kawin Ethayarajh.
\newblock How contextual are contextualized word representations? comparing the geometry of bert, elmo, and gpt-2 embeddings.
\newblock In {\em Proceedings of the 2019 conference on empirical methods in natural language processing and the 9th international joint conference on natural language processing (EMNLP-IJCNLP)}, pages 55--65, 2019.

\bibitem{gao2021simcse}
Tianyu Gao, Xingcheng Yao, and Danqi Chen.
\newblock Simcse: Simple contrastive learning of sentence embeddings.
\newblock In {\em Proceedings of the 2021 conference on empirical methods in natural language processing}, pages 6894--6910, 2021.

\bibitem{grattafiori2024llama}
Aaron Grattafiori, Abhimanyu Dubey, Abhinav Jauhri, Abhinav Pandey, Abhishek Kadian, Ahmad Al-Dahle, Aiesha Letman, Akhil Mathur, Alan Schelten, Alex Vaughan, et~al.
\newblock The llama 3 herd of models.
\newblock {\em arXiv preprint arXiv:2407.21783}, 2024.

\bibitem{gu2023mamba}
Albert Gu and Tri Dao.
\newblock Mamba: Linear-time sequence modeling with selective state spaces.
\newblock {\em arXiv preprint arXiv:2312.00752}, 2023.

\bibitem{guu2020retrieval}
Kelvin Guu, Kenton Lee, Zora Tung, Panupong Pasupat, and Mingwei Chang.
\newblock Retrieval augmented language model pre-training.
\newblock In {\em International conference on machine learning}, pages 3929--3938. PMLR, 2020.

\bibitem{hao2022iron}
Meng Hao, Hongwei Li, Hanxiao Chen, Pengzhi Xing, Guowen Xu, and Tianwei Zhang.
\newblock Iron: Private inference on transformers.
\newblock {\em Advances in neural information processing systems}, 35:15718--15731, 2022.

\bibitem{hyvarinen2000independent}
Aapo Hyv{\"a}rinen and Erkki Oja.
\newblock Independent component analysis: algorithms and applications.
\newblock {\em Neural networks}, 13(4-5):411--430, 2000.

\bibitem{izacard2021leveraging}
Gautier Izacard and Edouard Grave.
\newblock Leveraging passage retrieval with generative models for open domain question answering.
\newblock In {\em Proceedings of the 16th conference of the european chapter of the association for computational linguistics: main volume}, pages 874--880, 2021.

\bibitem{jacob2018quantization}
Benoit Jacob, Skirmantas Kligys, Bo~Chen, Menglong Zhu, Matthew Tang, Andrew Howard, Hartwig Adam, and Dmitry Kalenichenko.
\newblock Quantization and training of neural networks for efficient integer-arithmetic-only inference.
\newblock In {\em Proceedings of the IEEE conference on computer vision and pattern recognition}, pages 2704--2713, 2018.

\bibitem{johnson2019billion}
Jeff Johnson, Matthijs Douze, and Herv{\'e} J{\'e}gou.
\newblock Billion-scale similarity search with gpus.
\newblock {\em IEEE transactions on big data}, 7(3):535--547, 2019.

\bibitem{khandelwal2019generalization}
Urvashi Khandelwal, Omer Levy, Dan Jurafsky, Luke Zettlemoyer, and Mike Lewis.
\newblock Generalization through memorization: Nearest neighbor language models.
\newblock {\em arXiv preprint arXiv:1911.00172}, 2019.

\bibitem{knott2021crypten}
Brian Knott, Shobha Venkataraman, Awni Hannun, Shubho Sengupta, Mark Ibrahim, and Laurens van~der Maaten.
\newblock Crypten: Secure multi-party computation meets machine learning.
\newblock {\em Advances in Neural Information Processing Systems}, 34:4961--4973, 2021.

\bibitem{kwon2025vllm}
Woosuk Kwon.
\newblock {\em vLLM: An Efficient Inference Engine for Large Language Models}.
\newblock PhD thesis, UC Berkeley, 2025.

\bibitem{leviathan2023fast}
Yaniv Leviathan, Matan Kalman, and Yossi Matias.
\newblock Fast inference from transformers via speculative decoding.
\newblock In {\em International Conference on Machine Learning}, pages 19274--19286. PMLR, 2023.

\bibitem{lewis2020retrieval}
Patrick Lewis, Ethan Perez, Aleksandra Piktus, Fabio Petroni, Vladimir Karpukhin, Naman Goyal, Heinrich K{\"u}ttler, Mike Lewis, Wen-tau Yih, Tim Rockt{\"a}schel, et~al.
\newblock Retrieval-augmented generation for knowledge-intensive nlp tasks.
\newblock {\em Advances in neural information processing systems}, 33:9459--9474, 2020.

\bibitem{li2025collaborative}
Senyao Li, Haozhao Wang, Wenchao Xu, Rui Zhang, Song Guo, Jingling Yuan, Xian Zhong, Tianwei Zhang, and Ruixuan Li.
\newblock Collaborative inference and learning between edge slms and cloud llms: A survey of algorithms, execution, and open challenges.
\newblock {\em arXiv preprint arXiv:2507.16731}, 2025.

\bibitem{liu2026ministral}
Alexander~H Liu, Kartik Khandelwal, Sandeep Subramanian, Victor Jouault, Abhinav Rastogi, Adrien Sad{\'e}, Alan Jeffares, Albert Jiang, Alexandre Cahill, Alexandre Gavaudan, et~al.
\newblock Ministral 3.
\newblock {\em arXiv preprint arXiv:2601.08584}, 2026.

\bibitem{liu2023ring}
Hao Liu, Matei Zaharia, and Pieter Abbeel.
\newblock Ring attention with blockwise transformers for near-infinite context.
\newblock {\em arXiv preprint arXiv:2310.01889}, 2023.

\bibitem{luo2025shadow}
Zhifan Luo, Shuo Shao, Su~Zhang, Lijing Zhou, Yuke Hu, Chenxu Zhao, Zhihao Liu, and Zhan Qin.
\newblock Shadow in the cache: Unveiling and mitigating privacy risks of kv-cache in llm inference.
\newblock {\em arXiv preprint arXiv:2508.09442}, 2025.

\bibitem{min2020ambigqa}
Sewon Min, Julian Michael, Hannaneh Hajishirzi, and Luke Zettlemoyer.
\newblock Ambigqa: Answering ambiguous open-domain questions.
\newblock In {\em Proceedings of the 2020 conference on empirical methods in natural language processing (EMNLP)}, pages 5783--5797, 2020.

\bibitem{morris2023text}
John Morris, Volodymyr Kuleshov, Vitaly Shmatikov, and Alexander~M Rush.
\newblock Text embeddings reveal (almost) as much as text.
\newblock In {\em Proceedings of the 2023 Conference on Empirical Methods in Natural Language Processing}, pages 12448--12460, 2023.

\bibitem{ouyang2022training}
Long Ouyang, Jeffrey Wu, Xu~Jiang, Diogo Almeida, Carroll Wainwright, Pamela Mishkin, Chong Zhang, Sandhini Agarwal, Katarina Slama, Alex Ray, et~al.
\newblock Training language models to follow instructions with human feedback.
\newblock {\em Advances in neural information processing systems}, 35:27730--27744, 2022.

\bibitem{paillier1999public}
Pascal Paillier.
\newblock Public-key cryptosystems based on composite degree residuosity classes.
\newblock In {\em International conference on the theory and applications of cryptographic techniques}, pages 223--238. Springer, 1999.

\bibitem{pan2024unifying}
Shirui Pan, Linhao Luo, Yufei Wang, Chen Chen, Jiapu Wang, and Xindong Wu.
\newblock Unifying large language models and knowledge graphs: A roadmap.
\newblock {\em IEEE Transactions on Knowledge and Data Engineering}, 36(7):3580--3599, 2024.

\bibitem{pang2024bolt}
Qi~Pang, Jinhao Zhu, Helen M{\"o}llering, Wenting Zheng, and Thomas Schneider.
\newblock Bolt: Privacy-preserving, accurate and efficient inference for transformers.
\newblock In {\em 2024 IEEE Symposium on Security and Privacy (SP)}, pages 4753--4771. IEEE, 2024.

\bibitem{rajpurkar2016squad}
Pranav Rajpurkar, Jian Zhang, Konstantin Lopyrev, and Percy Liang.
\newblock Squad: 100,000+ questions for machine comprehension of text.
\newblock In {\em Proceedings of the 2016 conference on empirical methods in natural language processing}, pages 2383--2392, 2016.

\bibitem{sabt2015trusted}
Mohamed Sabt, Mohammed Achemlal, and Abdelmadjid Bouabdallah.
\newblock Trusted execution environment: What it is, and what it is not.
\newblock In {\em 2015 IEEE Trustcom/BigDataSE/Ispa}, volume~1, pages 57--64. IEEE, 2015.

\bibitem{singhal2023large}
Karan Singhal, Shekoofeh Azizi, Tao Tu, S~Sara Mahdavi, Jason Wei, Hyung~Won Chung, Nathan Scales, Ajay Tanwani, Heather Cole-Lewis, Stephen Pfohl, et~al.
\newblock Large language models encode clinical knowledge.
\newblock {\em Nature}, 620(7972):172--180, 2023.

\bibitem{sogaard2018limitations}
Anders S{\o}gaard, Sebastian Ruder, and Ivan Vuli{\'c}.
\newblock On the limitations of unsupervised bilingual dictionary induction.
\newblock In {\em Proceedings of the 56th Annual Meeting of the Association for Computational Linguistics (Volume 1: Long Papers)}, pages 778--788, 2018.

\bibitem{thomas2025hidden}
Rahul~Krishna Thomas, Louai Zahran, Erica Choi, Akilesh Potti, Micah Goldblum, and Arka Pal.
\newblock Hidden no more: Attacking and defending private third-party llm inference.
\newblock In {\em Forty-second International Conference on Machine Learning}, 2025.

\bibitem{trivedi2022musique}
Harsh Trivedi, Niranjan Balasubramanian, Tushar Khot, and Ashish Sabharwal.
\newblock Musique: Multihop questions via single-hop question composition.
\newblock {\em Transactions of the Association for Computational Linguistics}, 10:539--554, 2022.

\bibitem{vaswani2017attention}
Ashish Vaswani, Noam Shazeer, Niki Parmar, Jakob Uszkoreit, Llion Jones, Aidan~N Gomez, {\L}ukasz Kaiser, and Illia Polosukhin.
\newblock Attention is all you need.
\newblock {\em Advances in neural information processing systems}, 30, 2017.

\bibitem{wang2021milvus}
Jianguo Wang, Xiaomeng Yi, Rentong Guo, Hai Jin, Peng Xu, Shengjun Li, Xiangyu Wang, Xiangzhou Guo, Chengming Li, Xiaohai Xu, et~al.
\newblock Milvus: A purpose-built vector data management system.
\newblock In {\em Proceedings of the 2021 international conference on management of data}, pages 2614--2627, 2021.

\bibitem{xu2025survey}
Zhichao Xu, Fengran Mo, Zhiqi Huang, Crystina Zhang, Puxuan Yu, Bei Wang, Jimmy Lin, and Vivek Srikumar.
\newblock A survey of model architectures in information retrieval.
\newblock {\em arXiv preprint arXiv:2502.14822}, 2025.

\bibitem{yang2025qwen3}
An~Yang, Anfeng Li, Baosong Yang, Beichen Zhang, Binyuan Hui, Bo~Zheng, Bowen Yu, Chang Gao, Chengen Huang, Chenxu Lv, et~al.
\newblock Qwen3 technical report.
\newblock {\em arXiv preprint arXiv:2505.09388}, 2025.

\bibitem{yang2024parallelizing}
Songlin Yang, Bailin Wang, Yu~Zhang, Yikang Shen, and Yoon Kim.
\newblock Parallelizing linear transformers with the delta rule over sequence length.
\newblock {\em Advances in neural information processing systems}, 37:115491--115522, 2024.

\bibitem{yang2018hotpotqa}
Zhilin Yang, Peng Qi, Saizheng Zhang, Yoshua Bengio, William Cohen, Ruslan Salakhutdinov, and Christopher~D Manning.
\newblock Hotpotqa: A dataset for diverse, explainable multi-hop question answering.
\newblock In {\em Proceedings of the 2018 conference on empirical methods in natural language processing}, pages 2369--2380, 2018.

\bibitem{yates2021pretrained}
Andrew Yates, Rodrigo Nogueira, and Jimmy Lin.
\newblock Pretrained transformers for text ranking: Bert and beyond.
\newblock In {\em Proceedings of the 14th ACM International Conference on web search and data mining}, pages 1154--1156, 2021.

\bibitem{yuan2025scx}
Mu~Yuan, Lan Zhang, Liekang Zeng, Siyang Jiang, Bufang Yang, Di~Duan, and Guoliang Xing.
\newblock Scx: Stateless kv-cache encoding for cloud-scale confidential transformer serving.
\newblock In {\em Proceedings of the ACM SIGCOMM 2025 Conference}, pages 39--54, 2025.

\bibitem{zeng2025privacyrestore}
Ziqian Zeng, Jianwei Wang, Junyao Yang, Zhengdong Lu, Haoran Li, Huiping Zhuang, and Cen Chen.
\newblock Privacyrestore: Privacy-preserving inference in large language models via privacy removal and restoration.
\newblock In {\em Proceedings of the 63rd Annual Meeting of the Association for Computational Linguistics (Volume 1: Long Papers)}, pages 10821--10855, 2025.

\bibitem{zhang2024mgte}
Xin Zhang, Yanzhao Zhang, Dingkun Long, Wen Xie, Ziqi Dai, Jialong Tang, Huan Lin, Baosong Yang, Pengjun Xie, Fei Huang, et~al.
\newblock mgte: Generalized long-context text representation and reranking models for multilingual text retrieval.
\newblock In {\em Proceedings of the 2024 Conference on Empirical Methods in Natural Language Processing: Industry Track}, pages 1393--1412, 2024.

\bibitem{zhang2025siren}
Yue Zhang, Yafu Li, Leyang Cui, Deng Cai, Lemao Liu, Tingchen Fu, Xinting Huang, Enbo Zhao, Yu~Zhang, Yulong Chen, et~al.
\newblock Siren’s song in the ai ocean: A survey on hallucination in large language models.
\newblock {\em Computational Linguistics}, 51(4):1373--1418, 2025.

\bibitem{zhao2026retrieval}
Penghao Zhao, Hailin Zhang, Qinhan Yu, Zhengren Wang, Yunteng Geng, Fangcheng Fu, Ling Yang, Wentao Zhang, Jie Jiang, and Bin Cui.
\newblock Retrieval-augmented generation for ai-generated content: A survey.
\newblock {\em Data Science and Engineering}, pages 1--29, 2026.

\bibitem{zhao2023survey}
Wayne~Xin Zhao, Kun Zhou, Junyi Li, Tianyi Tang, Xiaolei Wang, Yupeng Hou, Yingqian Min, Beichen Zhang, Junjie Zhang, Zican Dong, et~al.
\newblock A survey of large language models.
\newblock {\em arXiv preprint arXiv:2303.18223}, 1(2):1--124, 2023.

\bibitem{zheng2024permllm}
Fei Zheng, Chaochao Chen, Zhongxuan Han, and Xiaolin Zheng.
\newblock Permllm: Private inference of large language models within 3 seconds under wan.
\newblock {\em arXiv preprint arXiv:2405.18744}, 2024.

\bibitem{zheng2025review}
Yue Zheng, Yuhao Chen, Bin Qian, Xiufang Shi, Yuanchao Shu, and Jiming Chen.
\newblock A review on edge large language models: Design, execution, and applications.
\newblock {\em ACM Computing Surveys}, 57(8):1--35, 2025.

\bibitem{zhong2021qmsum}
Ming Zhong, Da~Yin, Tao Yu, Ahmad Zaidi, Mutethia Mutuma, Rahul Jha, Ahmed Hassan, Asli Celikyilmaz, Yang Liu, Xipeng Qiu, et~al.
\newblock Qmsum: A new benchmark for query-based multi-domain meeting summarization.
\newblock In {\em Proceedings of the 2021 Conference of the North American Chapter of the Association for Computational Linguistics: Human Language Technologies}, pages 5905--5921, 2021.

\bibitem{zhong2022training}
Zexuan Zhong, Tao Lei, and Danqi Chen.
\newblock Training language models with memory augmentation.
\newblock In {\em Proceedings of the 2022 Conference on Empirical Methods in Natural Language Processing}, pages 5657--5673, 2022.

\end{thebibliography}

\clearpage
\appendix

\section{Numerical Stability Evaluation}\label{sec:appendix-numerical-stability}
We evaluate the numerical stability of using a random dense matrix for feature scrambling. Table~\ref{tab:attention-relative-error} presents the relative error between the scrambled attention outputs and the original plaintext results. We observe that scrambling with a random dense matrix introduces a relative error of approximately 12\%. In contrast, our proposed structured scrambling method ($S_1 P_1 H P_2 S_2$) significantly reduces this error to roughly 1.5\%, maintaining consistent stability across varying sequence lengths.

\begin{table}[ht]
\centering
\small
\setlength{\tabcolsep}{8pt}
\caption{Relative attention-output error.}
\label{tab:attention-relative-error}
\begin{tabular}{lccc}
\toprule
Method & Seq=128 & Seq=512 & Seq=2048 \\
\midrule
Dense Random Matrix & 12.61\% & 11.68\% & 12.62\% \\
$S_1 P_1 H P_2 S_2$ & 1.49\% & 1.52\% & 1.63\% \\
\bottomrule
\end{tabular}
\end{table}

\section{Dataset Descriptions}

\label{sec:appendix-datasets}

For accuracy evaluation, we select several prominent question-answering (QA) and summarization datasets equipped with reference contexts. These benchmarks cover a diverse range of domains, context lengths, and reasoning complexities:

\textbf{SQuAD~\cite{rajpurkar2016squad}.} SQuAD is a large-scale reading comprehension dataset comprising over 100,000 Wikipedia-based questions, where each answer is a continuous text span extracted directly from the provided passage.

\textbf{MuSiQue~\cite{trivedi2022musique}.} MuSiQue is a challenging multihop QA dataset constructed by composing single-hop questions. It minimizes dataset biases to ensure models perform genuine reasoning across multiple contexts rather than relying on shortcuts.

\textbf{QMSum~\cite{zhong2021qmsum}.} QMSum is a query-based meeting summarization benchmark containing annotated transcripts from various domains, designed to evaluate the extraction and summarization of relevant context from long-form conversations.

\textbf{HotpotQA~\cite{yang2018hotpotqa}.} HotpotQA is a diverse, explainable multi-hop QA dataset that requires models to reason across multiple Wikipedia documents and synthesize answers while providing sentence-level supporting facts.

\textbf{MS MARCO~\cite{bajaj2016ms}.} MS MARCO is a large-scale reading comprehension dataset built from anonymized Bing search queries, featuring real-world questions paired with human-generated answers derived from multiple web documents.

\textbf{AmbigQA~\cite{min2020ambigqa}.} AmbigQA focuses on ambiguous open-domain questions that yield multiple valid interpretations, evaluating a system's ability to identify ambiguity and generate comprehensive answers covering various plausible contexts.

\section{Overall Performance Details}\label{sec:appendix-overall-details}

\begin{table*}[t]
\centering
\small
\setlength{\tabcolsep}{4pt}
\caption{Performance evaluation detailed results.}
\label{tab:appendix-overall-details}
\resizebox{\textwidth}{!}{\begin{tabular}{llcccccccc}
\toprule
\multirow{2}{*}{Length} & \multirow{2}{*}{Method} & \multicolumn{4}{c}{\texttt{protected\_layers} = 2} & \multicolumn{4}{c}{\texttt{protected\_layers} = full} \\
\cmidrule(lr){3-6} \cmidrule(lr){7-10}
 & & TTFT (s) $\downarrow$ & Dec. TPS $\uparrow$ & Traff. (MiB) $\downarrow$ & Comm. Rounds $\downarrow$ & TTFT (s) $\downarrow$ & Dec. TPS $\uparrow$ & Traff. (MiB) $\downarrow$ & Comm. Rounds $\downarrow$ \\
\midrule
\multicolumn{10}{c}{\textbf{\texttt{Qwen/Qwen3-4B-Instruct-2507}}} \\
\midrule
\multirow{3}{*}{Short} & Ours & \textbf{0.25} & \textbf{36.15} & \textbf{156.47} & \textbf{75} & \textbf{3.65} & \textbf{9.91} & \textbf{328.01} & \textbf{1287} \\
 & PermLLM & 0.85 & 16.38 & 452.59 & 683 & 12.97 & 1.42 & 5135.92 & 12195 \\
 & SCX & 5.20 & 25.66 & 1418.02 & 628 & 5.20 & 4.86 & 1677.80 & 3878 \\
\midrule
\multirow{3}{*}{Medium} & Ours & \textbf{0.37} & \textbf{36.51} & \textbf{619.11} & \textbf{139} & \textbf{5.39} & \textbf{9.96} & \textbf{1210.85} & \textbf{2439} \\
 & PermLLM & 6.02 & 15.32 & 3650.17 & 1323 & 92.52 & 1.28 & 49936.56 & 23715 \\
 & SCX & 14.44 & 25.84 & 5494.15 & 820 & 15.20 & 4.84 & 6345.42 & 7334 \\
\midrule
\multirow{3}{*}{Long} & Ours & \textbf{1.05} & \textbf{36.79} & \textbf{2462.82} & \textbf{267} & \textbf{13.15} & \textbf{10.02} & \textbf{4639.35} & \textbf{4743} \\
 & PermLLM & 77.53 & 11.75 & 42196.11 & 2603 & 1341.59 & 0.86 & 640416.59 & 46755 \\
 & SCX & 52.82 & 26.22 & 21620.40 & 1204 & 55.41 & 4.83 & 24644.67 & 14246 \\
\midrule
\multicolumn{10}{c}{\textbf{\texttt{meta-llama/Llama-3.1-8B-Instruct}}} \\
\midrule
\multirow{3}{*}{Short} & Ours & \textbf{0.25} & \textbf{47.04} & \textbf{141.01} & \textbf{75} & \textbf{3.23} & \textbf{11.70} & \textbf{301.17} & \textbf{1147} \\
 & PermLLM & 0.86 & 18.18 & 439.03 & 683 & 11.93 & 1.62 & 4721.73 & 10851 \\
 & SCX & 4.86 & 30.97 & 1526.88 & 580 & 6.22 & 5.13 & 1791.25 & 3458 \\
\midrule
\multirow{3}{*}{Medium} & Ours & \textbf{0.42} & \textbf{46.99} & \textbf{555.65} & \textbf{139} & \textbf{4.83} & \textbf{11.79} & \textbf{1107.61} & \textbf{2171} \\
 & PermLLM & 5.93 & 16.94 & 3592.90 & 1323 & 89.23 & 1.46 & 45815.71 & 21091 \\
 & SCX & 17.44 & 28.67 & 5891.63 & 772 & 15.09 & 5.48 & 6752.25 & 6530 \\
\midrule
\multirow{3}{*}{Long} & Ours & \textbf{1.15} & \textbf{44.49} & \textbf{2207.37} & \textbf{267} & \textbf{12.00} & \textbf{11.85} & \textbf{4238.54} & \textbf{4219} \\
 & PermLLM & 77.51 & 12.74 & 41963.67 & 2603 & 1243.09 & 0.96 & 587393.74 & 41571 \\
 & SCX & 58.06 & 31.44 & 23159.13 & 1156 & 60.15 & 5.50 & 26202.25 & 12674 \\
\midrule
\multicolumn{10}{c}{\textbf{\texttt{mistralai/Ministral-3-14B-Instruct-2512-BF16}}} \\
\midrule
\multirow{3}{*}{Short} & Ours & \textbf{0.20} & \textbf{33.77} & \textbf{173.14} & \textbf{75} & \textbf{1.64} & \textbf{9.12} & \textbf{357.67} & \textbf{1427} \\
 & PermLLM & 0.93 & 19.68 & 466.64 & 558 & 14.73 & 1.29 & 5620.91 & 13539 \\
 & SCX & 9.09 & 24.45 & 2117.09 & 676 & 9.06 & 4.07 & 2406.17 & 4298 \\
\midrule
\multirow{3}{*}{Medium} & Ours & \textbf{0.43} & \textbf{33.05} & \textbf{683.78} & \textbf{139} & \textbf{3.51} & \textbf{9.23} & \textbf{1316.92} & \textbf{2707} \\
 & PermLLM & 6.07 & 14.62 & 3721.02 & 1323 & 106.02 & 1.52 & 53448.83 & 20139 \\
 & SCX & 19.91 & 24.56 & 8172.84 & 868 & 21.65 & 4.18 & 9083.67 & 8138 \\
\midrule
\multirow{3}{*}{Long} & Ours & \textbf{1.29} & \textbf{31.59} & \textbf{2719.49} & \textbf{267} & \textbf{12.51} & \textbf{9.17} & \textbf{5042.98} & \textbf{5267} \\
 & PermLLM & 76.65 & 11.23 & 42475.80 & 2603 & 1464.34 & 3.39 & 673616.17 & 12839 \\
 & SCX & 79.65 & 24.75 & 32134.34 & 1252 & 80.99 & 4.25 & 35278.67 & 15818 \\
\midrule
\multicolumn{10}{c}{\textbf{\texttt{openai/gpt-oss-20b}}} \\
\midrule
\multirow{3}{*}{Short} & Ours & \textbf{0.22} & \textbf{20.67} & \textbf{46.00} & \textbf{75} & \textbf{1.39} & \textbf{10.19} & \textbf{220.66} & \textbf{867} \\
 & PermLLM & 0.98 & 13.25 & 379.52 & 683 & 9.68 & 1.79 & 4224.67 & 8163 \\
 & SCX & 3.13 & 17.79 & 941.40 & 484 & 3.52 & 5.94 & 1181.29 & 2618 \\
\midrule
\multirow{3}{*}{Medium} & Ours & \textbf{0.44} & \textbf{20.66} & \textbf{160.84} & \textbf{139} & \textbf{2.45} & \textbf{10.28} & \textbf{804.73} & \textbf{1635} \\
 & PermLLM & 6.21 & 12.56 & 3295.09 & 1323 & 73.47 & 1.62 & 38419.13 & 15843 \\
 & SCX & 9.29 & 17.90 & 3643.90 & 676 & 10.13 & 5.98 & 4454.75 & 4922 \\
\midrule
\multirow{3}{*}{Long} & Ours & \textbf{1.35} & \textbf{21.88} & \textbf{611.33} & \textbf{140} & \textbf{7.72} & \textbf{13.54} & \textbf{3037.73} & \textbf{1647} \\
 & PermLLM & 77.91 & 10.24 & 40677.89 & 2603 & 966.51 & 1.16 & 483843.42 & 31203 \\
 & SCX & 36.42 & 18.06 & 14331.90 & 1060 & 39.13 & 6.01 & 17274.69 & 9530 \\
\bottomrule
\end{tabular}}
\end{table*}

Detailed performance data for the evaluations conducted in \S~\ref{sec:evaluation-throughput} are presented in Table~\ref{tab:appendix-overall-details}. We report the following evaluation metrics: Time to First Token (TTFT, in seconds), Decode Throughput (Dec. TPS, in tokens/s), Network Traffic (Traff., in MiB), and Communication Rounds . The optimal value for each metric is highlighted in bold. As demonstrated, our method consistently outperforms the baselines across all configurations and evaluation metrics.

\section{Benchmark Accuracy Details}\label{sec:appendix-qa-acc-details}

Detailed results for the benchmark evaluations discussed in \S~\ref{sec:evaulation-acc} are provided in Table~\ref{tab:qa-benchmark-results-details}. For the HotpotQA, MuSiQue, and MS MARCO datasets, we default to filtering the reference contexts using the supporting evidence (ground truth) provided by the datasets. This filtering mechanism simulates a practical RAG pipeline where a retriever isolates relevant passages from irrelevant distractors. For comprehensive comparison, we also evaluate the models by feeding all provided contexts (including distractors); these results are denoted with an \textit{unfiltered} suffix.

Regarding our accuracy metric, a response is considered correct if it contains the ground truth string as a substring. Consequently, the reported accuracy may occasionally exceed the ROUGE-L score. Note that QMSum is a meeting summarization dataset; since strict substring matching yields an accuracy of 0 across all models, we report the F1 score instead. Additionally, we conducted identical evaluations using the closed-source GPT-5.4 model, with results summarized in Table~\ref{tab:qa-gpt-results}. It is important to note that model performance on these benchmarks is highly sensitive to prompt design and text preprocessing. Therefore, the accuracy figures reported herein are intended solely for relative comparison within the scope of our experimental setup.

\begin{table*}[t]
\centering
\small
\setlength{\tabcolsep}{4pt}
\caption{Detailed Benchmark Performance under Distributed Attention and Feature Scrambling.}
\label{tab:qa-benchmark-results-details}
\begin{tabular}{llccc@{}p{15pt}@{}ccc@{}p{15pt}@{}ccc}
\toprule
\multirow{2}{*}{Model} & \multirow{2}{*}{Method} & \multicolumn{3}{c}{SQUAD} &  & \multicolumn{3}{c}{HOTPOTQA} &  & \multicolumn{3}{c}{MARCO} \\
\cmidrule(lr){3-5} \cmidrule(lr){7-9} \cmidrule(lr){11-13}
 & & Acc (\%) $\uparrow$ & Rouge-L $\uparrow$ & PPL $\downarrow$ &  & Acc (\%) $\uparrow$ & Rouge-L $\uparrow$ & PPL $\downarrow$ &  & Acc (\%) $\uparrow$ & Rouge-L $\uparrow$ & PPL $\downarrow$ \\
\midrule
\multirow{3}{*}{\shortstack{Llama 3.1\\8B}} & Baseline & \textbf{88.33} & \textbf{89.75} & 1.85 &  & \textbf{66.58} & \textbf{75.11} & 1.92 &  & 24.11 & 48.14 & 2.05 \\
 & +Distributed & 88.29 & 89.74 & 1.85 &  & 66.47 & 75.01 & 1.92 &  & 24.13 & 48.11 & 2.05 \\
 & +Scrambler & 88.30 & 89.72 & \textbf{1.85} &  & 66.50 & 75.00 & \textbf{1.92} &  & \textbf{24.18} & \textbf{48.16} & \textbf{2.05} \\
\midrule
\multirow{3}{*}{\shortstack{Qwen 3\\4B}} & Baseline & 89.21 & 88.22 & 6.43 &  & 69.63 & 79.30 & \textbf{5.94} &  & \textbf{24.25} & \textbf{46.34} & \textbf{7.01} \\
 & +Distributed & \textbf{89.23} & 88.23 & 6.43 &  & \textbf{69.68} & \textbf{79.38} & 5.97 &  & 24.11 & 46.25 & 7.02 \\
 & +Scrambler & 89.18 & \textbf{88.24} & \textbf{6.43} &  & 69.67 & 79.32 & 5.95 &  & 24.23 & 46.34 & 7.05 \\
\midrule
\multirow{3}{*}{\shortstack{Ministral 3\\14B}} & Baseline & \textbf{86.44} & \textbf{89.76} & 1.60 &  & 68.41 & 78.51 & 1.72 &  & 18.03 & 42.44 & \textbf{2.35} \\
 & +Distributed & 86.40 & 89.75 & 1.60 &  & 68.43 & \textbf{78.52} & 1.72 &  & \textbf{18.09} & \textbf{42.46} & 2.35 \\
 & +Scrambler & 86.41 & 89.74 & \textbf{1.60} &  & \textbf{68.47} & 78.50 & \textbf{1.72} &  & 18.02 & 42.35 & 2.35 \\
\midrule
\multirow{3}{*}{\shortstack{Qwen 2.5\\32B}} & Baseline & 93.72 & 90.94 & 2.35 &  & \textbf{72.99} & \textbf{80.14} & 2.69 &  & \textbf{22.25} & \textbf{46.75} & 5.50 \\
 & +Distributed & 93.75 & 91.00 & \textbf{2.35} &  & 72.82 & 80.06 & \textbf{2.69} &  & 22.00 & 46.56 & 5.51 \\
 & +Scrambler & \textbf{93.80} & \textbf{91.02} & 2.35 &  & 72.83 & 80.03 & 2.69 &  & 22.00 & 46.49 & \textbf{5.50} \\
\midrule
\multirow{2}{*}{Model} & \multirow{2}{*}{Method} & \multicolumn{3}{c}{AMBIGQA} &  & \multicolumn{3}{c}{MUSIQUE} &  & \multicolumn{3}{c}{QMSUM} \\
\cmidrule(lr){3-5} \cmidrule(lr){7-9} \cmidrule(lr){11-13}
 & & Acc (\%) $\uparrow$ & Rouge-L $\uparrow$ & PPL $\downarrow$ &  & Acc (\%) $\uparrow$ & Rouge-L $\uparrow$ & PPL $\downarrow$ &  & F1 (\%) $\uparrow$ & Rouge-L $\uparrow$ & PPL $\downarrow$ \\
\midrule
\multirow{3}{*}{\shortstack{Llama 3.1\\8B}} & Baseline & \textbf{56.79} & \textbf{63.78} & 3.92 &  & \textbf{59.33} & 62.63 & 1.98 &  & \textbf{30.02} & \textbf{23.08} & 11.61 \\
 & +Distributed & 56.59 & 63.64 & 3.92 &  & 59.21 & 62.47 & 1.98 &  & 29.97 & 23.07 & 11.61 \\
 & +Scrambler & 56.44 & 63.53 & \textbf{3.92} &  & 59.21 & \textbf{62.64} & \textbf{1.97} &  & 29.87 & 22.92 & \textbf{11.60} \\
\midrule
\multirow{3}{*}{\shortstack{Qwen 3\\4B}} & Baseline & \textbf{58.04} & \textbf{61.09} & 46.02 &  & 57.10 & 62.13 & \textbf{7.12} &  & \textbf{25.75} & 19.66 & 16.65 \\
 & +Distributed & \textbf{58.04} & 61.05 & 45.93 &  & \textbf{57.34} & \textbf{62.30} & 7.13 &  & 25.70 & \textbf{19.81} & \textbf{16.65} \\
 & +Scrambler & 57.79 & 61.02 & \textbf{45.40} &  & 57.10 & 62.03 & 7.16 &  & 25.65 & 19.60 & 16.69 \\
\midrule
\multirow{3}{*}{\shortstack{Ministral 3\\14B}} & Baseline & 59.24 & 64.26 & \textbf{4.09} &  & 65.70 & 66.80 & 1.91 &  & \textbf{22.59} & \textbf{17.73} & 16.49 \\
 & +Distributed & \textbf{59.29} & 64.22 & 4.10 &  & 65.58 & 66.60 & \textbf{1.91} &  & 22.52 & 17.65 & \textbf{16.46} \\
 & +Scrambler & \textbf{59.29} & \textbf{64.37} & 4.11 &  & \textbf{65.87} & \textbf{66.93} & 1.91 &  & 22.54 & 17.65 & 16.46 \\
\midrule
\multirow{3}{*}{\shortstack{Qwen 2.5\\32B}} & Baseline & \textbf{25.67} & \textbf{66.97} & \textbf{14.11} &  & \textbf{73.60} & \textbf{75.58} & 2.31 &  & 26.81 & 20.50 & \textbf{19.80} \\
 & +Distributed & 25.62 & 66.90 & 14.11 &  & 73.15 & 75.27 & 2.31 &  & 28.01 & 20.98 & 19.82 \\
 & +Scrambler & 25.62 & 66.72 & 14.13 &  & 72.98 & 75.06 & \textbf{2.31} &  & \textbf{28.09} & \textbf{21.09} & 19.83 \\
\midrule
\multirow{2}{*}{Model} & \multirow{2}{*}{Method} & \multicolumn{3}{c}{HOTPOTQA (Unfiltered)} &  & \multicolumn{3}{c}{MUSIQUE (Unfiltered)} &  & \multicolumn{3}{c}{MARCO (Unfiltered)} \\
\cmidrule(lr){3-5} \cmidrule(lr){7-9} \cmidrule(lr){11-13}
 & & Acc (\%) $\uparrow$ & Rouge-L $\uparrow$ & PPL $\downarrow$ &  & Acc (\%) $\uparrow$ & Rouge-L $\uparrow$ & PPL $\downarrow$ &  & Acc (\%) $\uparrow$ & Rouge-L $\uparrow$ & PPL $\downarrow$ \\
\midrule
\multirow{3}{*}{\shortstack{Llama 3.1\\8B}} & Baseline & 56.87 & 66.01 & \textbf{2.14} &  & \textbf{35.79} & 40.11 & \textbf{2.93} &  & 18.66 & 34.04 & 2.22 \\
 & +Distributed & \textbf{56.96} & \textbf{66.02} & 2.15 &  & \textbf{35.79} & \textbf{40.15} & 2.93 &  & \textbf{18.77} & \textbf{34.04} & 2.22 \\
 & +Scrambler & 56.87 & 65.92 & 2.15 &  & \textbf{35.79} & 40.00 & 2.93 &  & 18.66 & 34.03 & \textbf{2.22} \\
\midrule
\multirow{3}{*}{\shortstack{Qwen 3\\4B}} & Baseline & 61.01 & 69.59 & \textbf{11.62} &  & 30.41 & 35.57 & \textbf{15.72} &  & \textbf{17.58} & \textbf{29.43} & 12.85 \\
 & +Distributed & \textbf{61.08} & \textbf{69.61} & 11.63 &  & \textbf{30.49} & \textbf{35.70} & 15.73 &  & 17.48 & 29.42 & \textbf{12.84} \\
 & +Scrambler & 60.93 & 69.56 & 11.83 &  & 30.33 & 35.60 & 15.91 &  & 17.52 & 29.39 & 12.95 \\
\midrule
\multirow{3}{*}{\shortstack{Ministral 3\\14B}} & Baseline & 61.63 & 71.41 & \textbf{1.89} &  & \textbf{47.62} & \textbf{49.86} & 2.48 &  & 14.24 & 27.31 & 2.99 \\
 & +Distributed & \textbf{61.74} & \textbf{71.51} & 1.90 &  & 47.29 & 49.60 & 2.48 &  & \textbf{14.34} & \textbf{27.34} & \textbf{2.99} \\
 & +Scrambler & 61.66 & 71.39 & 1.89 &  & 47.37 & 49.78 & \textbf{2.48} &  & 14.25 & 27.30 & 2.99 \\
\midrule
\multirow{3}{*}{\shortstack{Qwen 2.5\\32B}} & Baseline & \textbf{66.08} & \textbf{73.95} & \textbf{2.95} &  & \textbf{47.54} & \textbf{52.80} & \textbf{3.54} &  & \textbf{17.08} & 32.64 & \textbf{7.19} \\
 & +Distributed & 65.56 & 73.44 & 2.96 &  & 46.83 & 52.23 & 3.55 &  & 16.85 & 32.68 & 7.19 \\
 & +Scrambler & 65.64 & 73.48 & 2.96 &  & 47.08 & 52.37 & 3.55 &  & 16.92 & \textbf{32.69} & 7.20 \\
\bottomrule
\end{tabular}
\end{table*}

\begin{table}[ht]
\centering
\small
\setlength{\tabcolsep}{4pt}
\caption{Benchmark Performance of GPT-5.4.}
\label{tab:qa-gpt-results}
\begin{tabular}{lccc}
\toprule
Dataset & Acc (\%) $\uparrow$ & F1 (\%) $\uparrow$ & Rouge-L $\uparrow$ \\
\midrule
SQUAD & 94.24 & 91.98 & 91.79 \\
HOTPOTQA & 76.41 & 84.19 & 84.66 \\
MARCO & 21.28 & 45.32 & 46.86 \\
AMBIGQA & 65.93 & 69.50 & 70.23 \\
MUSIQUE & 82.25 & 83.39 & 83.49 \\
QMSUM & 0.00 & 24.38 & 17.89 \\
HOTPOTQA (Unfiltered) & 75.21 & 83.09 & 83.54 \\
MUSIQUE (Unfiltered) & 78.03 & 79.59 & 79.97 \\
MARCO (Unfiltered) & 16.94 & 29.34 & 30.31 \\
\bottomrule
\end{tabular}
\end{table}

\section{Evaluation of Quantization Effects}\label{sec:appendix-quantization}

Table~\ref{tab:qa-six-dataset-quant-delta} illustrates the utility degradation resulting from 8-bit quantization of LLM intermediate states, as discussed in \S~\ref{sec:evaluation-quantization}. For these experiments, we evaluated only the first 128 samples from each dataset. As noted in the main text, naively applying quantization introduces significant performance penalties. However, deliberately removing the outermost scaling matrix $S_2$ effectively mitigates these quantization-induced losses. Furthermore, our evaluation currently employs a rudimentary affine min-max quantization scheme; integrating more advanced quantization techniques could potentially yield even better accuracy-efficiency trade-offs in future implementations.

\begin{table*}[t]
\centering
\small
\setlength{\tabcolsep}{4pt}
\caption{Benchmark accuracy degradation introduced by different scrambling methods under 8-bit quantization.}
\label{tab:qa-six-dataset-quant-delta}
\begin{tabular}{llccc@{}p{15pt}@{}ccc@{}p{15pt}@{}ccc}
\toprule
\multirow{2}{*}{Model} & \multirow{2}{*}{Method} & \multicolumn{3}{c}{SQUAD} &  & \multicolumn{3}{c}{HOTPOTQA} &  & \multicolumn{3}{c}{MARCO} \\
\cmidrule(lr){3-5} \cmidrule(lr){7-9} \cmidrule(lr){11-13}
 & & $\Delta$Acc & $\Delta$Rouge-L & $\Delta$PPL &  & $\Delta$Acc & $\Delta$Rouge-L & $\Delta$PPL &  & $\Delta$Acc & $\Delta$Rouge-L & $\Delta$PPL \\
\midrule
\multirow{2}{*}{\shortstack{Llama 3.1\\8B}} & S$_1$ \& S$_2$ & -29.69 & -31.12 & +0.49 &  & -39.84 & -37.06 & +3.09 &  & -12.50 & -28.12 & +0.84 \\
 & S$_1$ only & \textbf{+0.00} & \textbf{-0.39} & \textbf{+0.05} &  & \textbf{+2.34} & \textbf{+2.14} & \textbf{+0.05} &  & \textbf{-1.56} & \textbf{-0.02} & \textbf{+0.01} \\
\midrule
\multirow{2}{*}{\shortstack{Qwen 3\\4B}} & S$_1$ \& S$_2$ & -20.31 & -25.49 & \textbf{-0.51} &  & -27.34 & -36.31 & +2.75 &  & -8.59 & -22.27 & \textbf{-2.12} \\
 & S$_1$ only & \textbf{+0.78} & \textbf{-1.95} & -0.47 &  & \textbf{-0.78} & \textbf{-1.01} & \textbf{-0.72} &  & \textbf{+0.00} & \textbf{+0.97} & -0.12 \\
\midrule
\multirow{2}{*}{\shortstack{Ministral 3\\14B}} & S$_1$ \& S$_2$ & -3.13 & -1.03 & +0.03 &  & -2.34 & -3.33 & +0.09 &  & \textbf{-1.56} & -2.52 & +0.08 \\
 & S$_1$ only & \textbf{+0.00} & \textbf{+0.00} & \textbf{+0.00} &  & \textbf{+0.78} & \textbf{+0.78} & \textbf{+0.00} &  & \textbf{-1.56} & \textbf{-2.22} & \textbf{-0.01} \\
\midrule
\multirow{2}{*}{\shortstack{Qwen 2.5\\32B}} & S$_1$ \& S$_2$ & -4.69 & -3.31 & +0.30 &  & -3.13 & -0.82 & +1.03 &  & +0.00 & -5.26 & +0.04 \\
 & S$_1$ only & \textbf{+0.00} & \textbf{+0.65} & \textbf{-0.01} &  & \textbf{+0.00} & \textbf{-0.13} & \textbf{-0.04} &  & \textbf{+0.78} & \textbf{+0.09} & \textbf{-0.07} \\
\midrule
\multirow{2}{*}{Model} & \multirow{2}{*}{Method} & \multicolumn{3}{c}{AMBIGQA} &  & \multicolumn{3}{c}{MUSIQUE} &  & \multicolumn{3}{c}{QMSUM} \\
\cmidrule(lr){3-5} \cmidrule(lr){7-9} \cmidrule(lr){11-13}
 & & $\Delta$Acc & $\Delta$Rouge-L & $\Delta$PPL &  & $\Delta$Acc & $\Delta$Rouge-L & $\Delta$PPL &  & $\Delta$F1 & $\Delta$Rouge-L & $\Delta$PPL \\
\midrule
\multirow{2}{*}{\shortstack{Llama 3.1\\8B}} & S$_1$ \& S$_2$ & -39.06 & -57.97 & +83.23 &  & -43.75 & -38.78 & +2.55 &  & -21.53 & -11.94 & +71.53 \\
 & S$_1$ only & \textbf{+0.78} & \textbf{+1.35} & \textbf{+0.06} &  & \textbf{+0.78} & \textbf{-0.36} & \textbf{+0.03} &  & \textbf{-0.64} & \textbf{-0.88} & \textbf{+0.01} \\
\midrule
\multirow{2}{*}{\shortstack{Qwen 3\\4B}} & S$_1$ \& S$_2$ & -31.25 & -49.15 & +50.30 &  & -35.94 & -34.60 & +1.75 &  & -13.59 & -5.99 & +37.00 \\
 & S$_1$ only & \textbf{-1.56} & \textbf{-1.20} & \textbf{-7.22} &  & \textbf{-5.47} & \textbf{-3.45} & \textbf{-0.48} &  & \textbf{-0.27} & \textbf{-0.05} & \textbf{-0.25} \\
\midrule
\multirow{2}{*}{\shortstack{Ministral 3\\14B}} & S$_1$ \& S$_2$ & -7.03 & -9.58 & \textbf{-0.02} &  & -10.16 & -6.90 & +0.25 &  & -1.67 & -0.94 & +0.85 \\
 & S$_1$ only & \textbf{-0.78} & \textbf{-1.04} & +0.01 &  & \textbf{+0.00} & \textbf{+0.52} & \textbf{+0.01} &  & \textbf{-0.05} & \textbf{+0.13} & \textbf{+0.06} \\
\midrule
\multirow{2}{*}{\shortstack{Qwen 2.5\\32B}} & S$_1$ \& S$_2$ & -10.94 & -16.86 & +10.81 &  & -14.06 & -9.56 & +1.79 &  & -12.06 & -5.82 & +35.23 \\
 & S$_1$ only & \textbf{+0.78} & \textbf{+0.33} & \textbf{-0.53} &  & \textbf{+0.00} & \textbf{+0.39} & \textbf{-0.04} &  & \textbf{+0.22} & \textbf{+0.13} & \textbf{-0.29} \\
\bottomrule
\end{tabular}
\end{table*}

\end{document}